\documentclass{pasj01}
\Received{$\langle$reception date$\rangle$}
\Accepted{$\langle$acception date$\rangle$}
\Published{$\langle$publication date$\rangle$}

\newcommand{{\maxi}}{{\it MAXI}}
\newcommand{{\swift}}{{\it Swift}}
\newcommand{{\nustar}}{{\it NuSTAR}}
\newcommand{{\nicer}}{{\it NICER}}

\newcommand{\Tin}{T_\mathrm{in}}

\newcommand{\noprint}[1]{}

\usepackage{url}
\usepackage{color}
\usepackage{multirow}
\usepackage{threeparttable}
\usepackage{natbib}
\usepackage{tabularx}
\usepackage[switch,mathlines]{lineno} 

\begin{document}
\title{Evolution of Accretion Disk Structure of the Black Hole X-ray
 Binary MAXI J1820$+$070 during the Rebrightening Phase}

\author{Tomohiro \textsc{Yoshitake}\altaffilmark{1}}%

\altaffiltext{1}{Department of Astronomy, Kyoto University, Kitashirakawa-Oiwake-cho, Sakyo-ku, Kyoto, Kyoto 606-8502, Japan}
\email{yoshitake@kusastro.kyoto-u.ac.jp}

\author{Megumi \textsc{Shidatsu}\altaffilmark{2}}

\altaffiltext{2}{Department of Physics, Ehime University, 
2-5, Bunkyocho, Matsuyama, Ehime 790-8577, Japan}

\author{Yoshihiro \textsc{Ueda}\altaffilmark{1}}
\author{Daisaku \textsc{Nogami}\altaffilmark{1}}
\author{Katsuhiro L. \textsc{Murata}\altaffilmark{3}}
\author{Narikazu \textsc{Higuchi}\altaffilmark{4}}
\author{Keisuke \textsc{Isogai\altaffilmark{3, 5}}}
\author{Hiroyuki \textsc{Maehara}\altaffilmark{3,4}}

\author{Shin \textsc{Mineshige}\altaffilmark{1}}
\author{Hitoshi \textsc{Negoro}\altaffilmark{6}}
\author{Nobuyuki \textsc{Kawai}\altaffilmark{4}}
\author{Yoichi \textsc{Yatsu}\altaffilmark{4}}
\author{Mahito \textsc{Sasada}\altaffilmark{4}}
\author{Ichiro \textsc{Takahashi}\altaffilmark{4}}
\author{Masafumi \textsc{Niwano}\altaffilmark{4}}

\altaffiltext{3}{Okayama Observatory, Kyoto University, 3037-5 Honjo, Kamogatacho, Asakuchi, Okayama 719-0232, Japan}
\altaffiltext{4}{Department of Physics, Tokyo Institute of Technology, 2-12-1 Ookayama, Meguro-ku, Tokyo 152-8551, Japan}
\altaffiltext{5}{Department of Multi-Disciplinary Sciences, Graduate School of Arts and Sciences, The University of Tokyo, 3-8-1 Komaba, Meguro, Tokyo 153-8902, Japan}
\altaffiltext{6}{Department of Physics, Nihon University, 1-8-14 Kanda-Surugadai, Chiyoda-ku, Tokyo 101-8308, Japan}

\author{Tomoki \textsc{Saito}\altaffilmark{7}}
\author{Masaki \textsc{Takayama}\altaffilmark{7}}
\altaffiltext{7}{Nishi-Harima Astronomical Observatory, Center for Astronomy, University of Hyogo, 407-2 Nishigaichi, Sayo-cho, Sayo, Hyogo 679-5313, Japan}
\author{Yumiko \textsc{Oasa}\altaffilmark{8,9}}
\altaffiltext{8}{Faculty of education, Saitama University, Simo-okubo 255, Sakura-ku, Saitama 338-8570, Japan}
\altaffiltext{9}{Graduate School of Science and Engineering, Saitama University, Simo-okubo 255, Sakura-ku, Saitama 338-8570, Japan}
\author{Takuya \textsc{Takarada}\altaffilmark{9,10}}
\altaffiltext{10}{Astrobiology Center, NINS, 2-21-1 Osawa, Mitaka, Tokyo 181-8588, Japan}
\author{Takumi \textsc{Shigeyoshi}\altaffilmark{8}}
\author{OISTER Collaboration}

\KeyWords{X-rays: individual (MAXI J1820$+$070) --- X-rays: binaries --- accretion, accretion disks --- black hole physics}

\maketitle

\begin{abstract}
To understand the evolution of global accretion disk structure in the ``rebrightening'' phase of MAXI J1820$+$070, we perform a
comprehensive analysis of its near infrared/optical/UV to X-ray
spectral energy distribution (SED) utilizing data obtained by OISTER, Las Cumbres Observatory (LCO), Swift, NICER, and 
NuSTAR in 2019.
Optical spectra observed with Seimei telescope in 2019 and 2020 are also analyzed.
On the basis of the optical and X-ray light curves and their flux ratios, we divide the whole phase into 3 periods, Periods I (flux rise), II (decay), and III (dim).
In the first 2 periods, the source stayed in the low/hard state (LHS),
where the X-ray (0.3--30 keV) and optical/UV SED can be both fitted with
power-law models.
We interpret that the X-ray emission arises from hot corona via Comptonization, whereas the
optical/UV flux is dominated by
synchrotron radiation from 
the jets, with a partial contribution from the irradiated disk.
The optical/UV power-law component 
smoothly connects to a simultaneous radio flux, supporting its jet origin. Balmer line profiles in the optical spectra
indicate that the inner radius of an irradiated disk slightly decreased from
$\sim 2\times 10^5 r_{\rm g}$ (Period I) to $\sim 1\times 10^5 r_{\rm g}$ (Period II),
where $r_{\rm g}$ is the gravitational radius, implying a change of the hot corona geometry.
In Period III, the SED can be reproduced by an advection-dominated accretion flow and jet
emission. However, the double-peaked H$\alpha$ emission line indicates that a cool disk remained at large radii.
\end{abstract}

\section{Introduction} \label{sec:intro}

Many Galactic black hole X-ray binaries (BHXB) have a transient nature. They suddenly exhibit outbursts and increase their X-ray luminosity by several orders of magnitude (e.g., \citealt{tanaka1996}; \citealt{tetarenko2016}). These large outbursts are often followed by one or more rebrightenings (or ``mini-outburts'') whose peak fluxes are usually lower than the main outbursts (\citealt{MunozDarias2017}; \citealt{cuneo2020}; \citealt{zhang2019}; \citealt{saikia2023}).
The X-rays from BHXBs are produced by the release of the gravitational energy of the accreted gas, and the luminosity 
is considered to reflect the mass accretion rate. Therefore, monitoring BHXBs during their outburst/rebrightening periods gives us insight on how the black hole accretion 
disk evolves with the mass accretion rate. 

Intensive observations, mainly in the X-ray band, have been revealing the accretion disk structure at high luminosities. 
After the outburst rise, BHXBs show the state transition from the low/hard state (LHS) to the high/soft state (HSS), where a hard, power-law shaped X-ray spectrum changes to 
a thermal spectrum dominating the soft X-ray band. 
This transition is generally explained by the change in the inner disk structure (see e.g., \citealt{don07}). In the LHS, the standard disk \citep{sha73}, which is an optically thick and geometrically thin accretion disk, is replaced by a hot, radiatively inefficient inner flow (RIAF) in the vicinity of the black hole (e.g., \citealt{mks08}), while in the HSS, the standard disk 
extends down to the innermost stable circular orbit (e.g., \citealt{ebs93}). 
However, studies at low luminosities, 
especially below an X-ray luminosity of 
$L_{\mathrm X} \sim 10^{36}$ erg s$^{-1}$, is limited so far.  
Although in theory, the outburst evolution of the disk 
is generally explained by the disk instability model 
\citep{mineshige1989, lasota2001},  
observational studies are limited to answer questions such 
as when the disk instability occurs at the outer disk 
region and how the standard disk extends inwards, replacing 
the RIAF, with increasing mass accretion rate.

Multi-wavelength observations are key to understand 
the whole picture of the accretion disk 
structure and its evolution, because the disks 
in BHXBs emit photons not only in X-rays but also different wavelengths. They mainly emit X-ray photons in the vicinity 
of the black hole, while the outer regions of the disks predominantly produce ultraviolet (UV), optical, and near-infrared (IR) emission. 
Previous studies based on multi-wavelength spectral energy distribution (SED) suggest that the outer disk regions are heated by significant X-ray irradiation, resulting in enhancement of optical and near-IR fluxes (\citealt{gie08}; \citealt{gie09}). 
Strong X-ray irradiation can develop a hot, ionized atmosphere 
above the disk (e.g., \citealt{2002ApJ...581.1297J}) and produce 
emission lines in the UV, optical, and near-IR bands. The lines 
usually show a double-peaked profile, due to the Keplerian motion 
of the gas in the line emitting region \citep{horne1986}. 
It can also drive powerful disk winds \citep{begelman1983}. 
Indeed, previous X-ray, optical, and near-IR spectroscopy of BHXBs detected blueshifted absorption lines (or P Cygni profiles), suggestive 
of the presence of disk winds (e.g., \citealt{ueda1998}; \citealt{MunozDarias2016}; \citealt{Sanchez2020}). These results demonstrate that multi-wavelength studies are essential for full understanding of black hole accretion and outflows. 

MAXI J1820 is a Galactic BHXB discovered with Monitor of All-sky X-ray Image (MAXI; \citealt{mat09}) in 2018 \citep{kwm18, sdt18, tuk18}. After the main outburst in 2018 with state transitions between the HSS and LHS (e.g., \citealt{sdt19}), it showed several rebrightening events, 
during which no state transitions to the HSS were reported. 
These outburst and rebrightening events were intensively observed at various wavelengths and 
using various methods, including multi-wavelength SED modeling (\citealt{2021ApJ...910...21R}; \citealt{ozbey2022}; \citealt{Echiburu-Trujillo2023}) and spectroscopy of optical emission and absorption lines (\citealt{MunozDarias2019}; \citealt{Sanchez2020}; \citealt{sai2021}; \citealt{kolijonen2023}; \citealt{tetarenko2023}).
In \cite{yoshitake2022}, we used a near-IR to X-ray SED and optical spectrum taken at an X-ray luminosity of $\sim 10^{33}$ erg s$^{-1}$ in the first rebrightening in 2019, and found that the emission from the advection-dominant accretion flow (ADAF) is dominant over the entire bands at this epoch.

In this article, we compile the multi-wavelength SEDs of MAXI J1820 obtained at different luminosities in the 2019 and 2020 rebrightenings, and investigate the evolution of the disk structure at low mass accretion rates, with an X-ray luminosity range of $\sim 6 \times 10^{32}$ erg s$^{-1} < L_{\mathrm X} \lesssim 6 \times 10^{35}$ erg s$^{-1}$ (which corresponds to an Eddington ratio of $\sim 10^{-6} < L_{\mathrm X}/L_{\mathrm Edd} \lesssim 10^{-3}$). 
In addition, we use optical spectra obtained during these rebrightenings, to identify emission and absorption lines and study the line profiles and their variation with X-ray luminosity. In this work, we assume a distance of $D = 3$ kpc \citep{gan19}, a black hole mass of $M = $ 7--8 $M_\odot$, and an inclination angle of $i = 69^\circ$--77$^\circ$ \citep{tor19} for MAXI J1820.

\section{Observations and Data Reduction}
\label{sec:obs}

We collected multi-wavelength data of MAXI J1820 obtained with various observatories in the first rebrightening period in 2019. 
We used X-ray data taken by the {\it Swift}/XRT, {\it NuSTAR} and {\it NICER}, UV data from the {\it Swift}/UVOT, and optical and near-IR photometric data from telescopes of Las Cumbres Observatory (LCO) and those in the Optical and Infrared Synergetic Telescopes for Education and Research (OISTER) collaboration. 
We also performed optical spectroscopic observations occasionally with {\it Seimei} telescope in the 2019 and 2020 rebrightenings.  Table~\ref{tab:observation_x} and Table~\ref{tab:observation_opt} give a log of the X-ray and UV observations and that of the optical and near-IR observations, respectively.
Note that there are more LCO, {\it Swift}, and {\it NICER} observations 
than those listed in the tables, but we omitted the ones 
without sufficient simultaneous multi-wavelength coverage 
(see Section~\ref{subsec:SEDana} for more detail). 

\begin{table*}
  \centering
\begin{threeparttable}
    \tbl{X-Ray and UV ({\it Swift}/UVOT) observation log}{
    \begin{tabularx}{\linewidth}{lccc}
  \hline \hline
  {\bf Observatory/Instrument (Filters))} & {\bf Date} & {\bf Exposure (ks)}& {\bf Observation ID} \\ \hline
  {\it Swift}/XRT & 2019 Mar 16 & 1 & 10627140 \\
  {\it Swift}/UVOT (UVM2) & 2019 Mar 16 & 1 & 10627140 \\
  {\it Swift}/XRT & 2019 Mar 17 & 2 & 10627141 \\
  {\it Swift}/UVOT (U, V, UVM2, UVW1, UVW2)  & 2019 Mar 17 & 0.1 each & 10627141 \\
  {\it Swift}/XRT & 2019 Mar 24 & 2 & 10627148 \\
  {\it Swift}/UVOT  (U, V,  UVW1) & 2019 Mar 24 & 0.1 each & 10627148 \\
  {\it Swift}/UVOT  (UVM2,  UVW2) & 2019 Mar 24 & 0.3 each & 10627148 \\
  {\it Swift}/XRT &  2019 Mar 29 & 0.6 & 10627150 \\
  {\it Swift}/XRT &  2019 Apr 10 & 0.7 & 10627158 \\
  {\it NICER}/XTI & 2019 Apr 30 & 2.6 & 2200120329 \\
  {\it NuSTAR}/FPMA, FPMB & 2019 May 1 & 42 (FPMA), 41(FPMB) & 90501320002 \\
  {\it Swift}/XRT & 2019 May 10--13 & 4 & 10627169/171/172/173 \\
  {\it Swift}/UVOT (U, UVM2, UVW1, UVW2) & 2019 May 11 & 0.1 each &10627171 \\
  \hline
  \end{tabularx}}
  \label{tab:observation_x}
\end{threeparttable}
\end{table*}

\begin{table*}
\begin{threeparttable}
  \centering
    \tbl{Optical and near-IR observation log}{
    \begin{tabularx}{\linewidth}{lccc}
  \hline \hline
  {\bf Observatory/Instrument (Filters))} & {\bf Date} & {\bf Exposure (ks)}& {\bf Observation ID} \\ \hline
  OISTER (g) & 2019 Mar 16 & &\\
  LCO, ogg\tnote{$\dagger$}  (iVR) & 2019 Mar 16 & 0.2 each & 65, 66, 67 \\
  OISTER (g) & 2019 Mar 17 & &\\
  LCO, cpt\tnote{$\dagger$} (gr) & 2019 Mar 29 & 0.2 each & 171, 173\\
  LCO, cpt (y) & 2019 Mar 29 & 0.3 & 172 \\
  LCO, elp\tnote{$\dagger$} (ri) & 2019 Apr 9 & 0.06 each & 183,186 \\
  LCO, elp (y) & 2019 Apr 9 & 0.1 & 185 \\
  LCO, cpt (gri) & 2019 May 1 & 0.06 each & 214,215,217\\
  OISTER (JHK) & 2019 May 1 & &\\
  OISTER (grizJHK) & 2019 May 10 & &\\
  {\it Seimei}/KOOLS-IFU & 2019 May 11 & 0.6 ($\times$15 frames)& \\
  {\it Seimei}/KOOLS-IFU & 2020 Feb 23 & 0.3 ($\times$16 frames)& \\
  {\it Seimei}/KOOLS-IFU & 2020 Mar 18 & 0.3 ($\times$ 16 frames)& \\
  \hline
  \end{tabularx}}
  \label{tab:observation_opt}
  \smallskip
  \begin{tablenotes}[normal]
   \footnotesize
   \small
    \item[$\dagger$] %
      The site codes ogg, cpt, and elp indicate Haleakala Observatory, South African Astronomical Observatory, and McDonald Observatory.
    \end{tablenotes}
\end{threeparttable}
\end{table*}

\subsection{\swift}

The time-averaged {\it Swift}/XRT spectrum in the individual OBSID were obtained via the XRT on-demand web interface\footnote{\url{https://www.swift.ac.uk/user_objects/}}. 
The XRT data that we used were obtained either of the two CCD 
readout mode: the Photon Counting (PC) mode, which provides 2 
dimensional spatial information, or the Windowed Timing (WT) mode, 
which retains only 1-dimensional information \citep{bur05}.
To improve the statistics, we combined the data taken in the  
2019 May 10--13 observations (OBSIDs$=$00010627169, 00010627171, 00010627172, and 00010627173), which correspond to the end of 
the rebrightening. 

The {\it Swift}/UVOT data were reduced with HEAsoft version 6.31.1 
and the latest {\it Swift}/UVOT Calibration Database (CALDB) 
as of 2021 November. We adopted the cleaned sky-coordinate 
images of the individual filters: UVW1, UVW2, UVM2, U, and V 
bands, and performed aperture photometry using the UVOT tool {\tt uvot2pha} included in HEAsoft. Here, we defined the source and background regions as circles with a $10"$   
radius centered at the source position and in a blank-sky area, respectively. 

\subsection{\nustar}

The {\it NuSTAR} data were downloaded from the HEASARC archive\footnote{\url{https://heasarc.gsfc.nasa.gov/docs/archive.html}} and reduced with {\tt nupipeline} version 0.4.9 in the {\it NuSTAR} Data Analysis Software ({\tt nustardas}), utilizing the calibration database (CALDB) version 20191219. The source and background extraction regions were defined as circular regions with a radius of $100"$ centered on the target position and in a nearby source-free region on the same chip as that of the source, respectively. 
The spectra of the individual focal plane modules (FPMA and FPMB) were produced through {\tt nuproducts} with standard settings for 
a point source described in the NuSTAR quickstart guide\footnote{\url{https://heasarc.gsfc.nasa.gov/docs/nustar/analysis/nustar_quickstart_guide.pdf}}. 
We combined the spectra from the FPMA and FPMB with {\tt addascaspec}.

\subsection{\nicer}
The {\it NICER} data were downloaded from the HEASARC archive. 
We performed pipeline processing using {\tt nicerl2} 
included in HEAsoft, referring to the {\it NICER} CALDB 
version 20190516. 
The background spectrum was build with the 
background estimator tool {\tt nibackgen3C50} and 
the response matrix file and the ancillary response 
file were created with the {\tt nicerrmf} and 
{\tt nicerarf}, respectively, all of which 
are included in HEAsoft.

\subsection{OISTER}

Optical and near-IR multi-band photometric data 
were obtained by the OISTER collaboration, which 
is formed with many small-to-medium size telescopes 
operated by Japanese universities and observatories, 
including  Multicolor Imaging Telescopes for Survey 
and Monstrous Explosions (MITSuME) at the Akeno Observatory \citep{kot05, mitsumeAkeno2007, mitsumeAkeno2008} and at the Okayama Observatory \citep{mitsumeOkayama2010}, 55 cm SaCRA telescope 
at Saitama University \citep{sacraMusashi2020}, and the 2.0 m 
Nayuta telescope at the Nishi-Harima Astronomical Observatory \citep{nic2011, nic2013}. 

Using IRAF \citep{bar94, bar95}, we conducted standard data 
reduction procedures including bias and dark subtraction, 
flat fielding, bad pixel masking, and photometry. 
The magnitude calibration was performed with nearby reference 
stars, whose magnitudes were taken from Pan-STARRS1 Surveys
\citep{2016arX161205560C} for the optical data, and the 
Two MicronAll Sky Survey Point Source Catalog \citep{2003tmc..book.....C} for
the near-IR data. Their typical errors were 
estimated to be $\sim$10\% of the resultant flux density. 

\begin{figure}[htbp]
  \begin{center}
     \includegraphics[width=80mm]{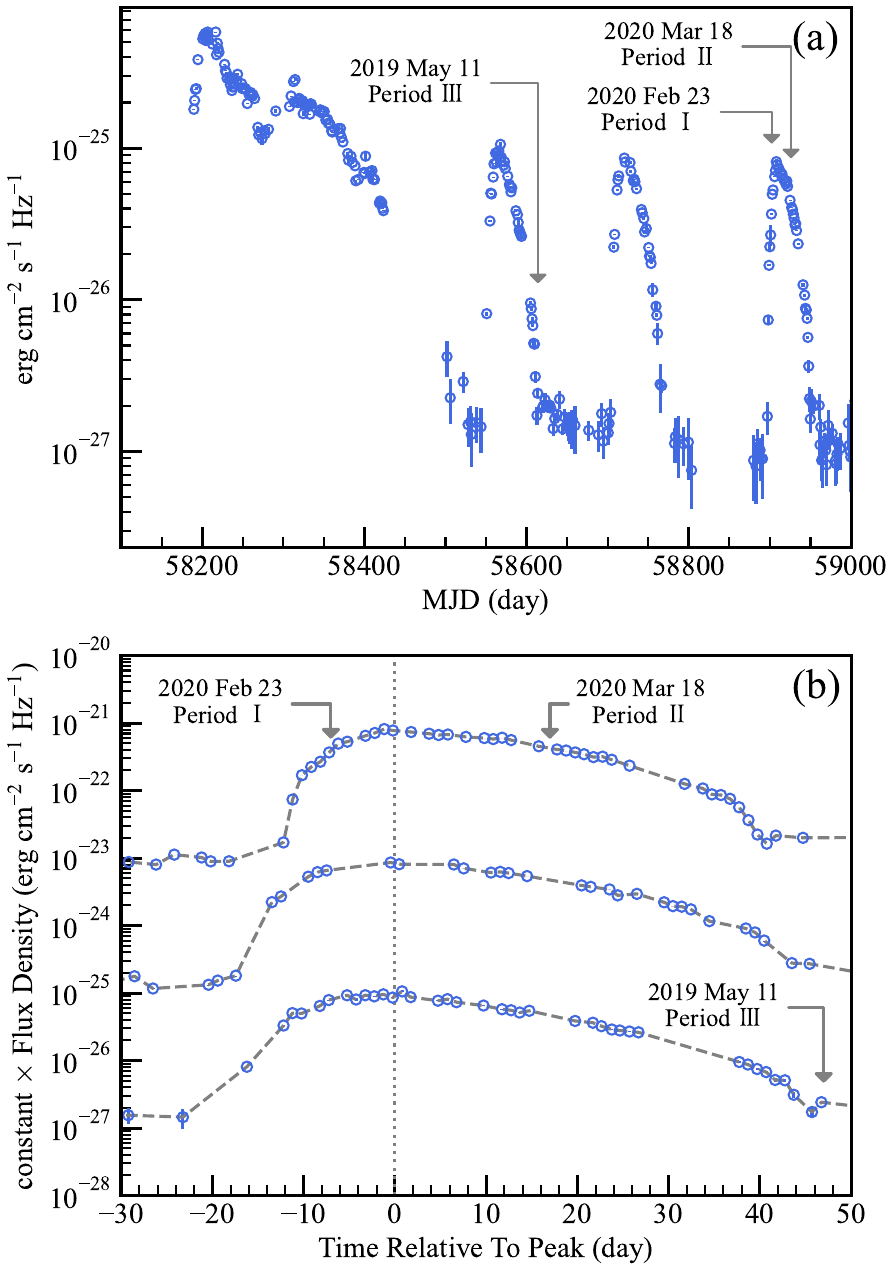} 
  \end{center}
  \caption{(a) Optical (g'-band) light curve of MAXI J1820 from the OISTER collaboration (Higuchi et al. in preparation).  
  MJD 58200 corresponds to 2018 March 23. (b) The g'-band light curves in the three rebrightening phases, in which the values on the horizontal axis are adjusted with the flux peaks (MJD 58567, 58721, and 58909 for the first, second, and third rebrightenings, respectively). The flux densities in the second and third rebrightenings are multiplied by $10^2$ and $10^4$, respectively, for plotting purposes. The {\it Seimei} observations are indicated in each panel.}
  \label{fig:lc_opt}
\end{figure}

\subsection{LCO}
We used archival LCO data of MAXI J1820 in the first 
rebrighening period. From the science data archive\footnote{https://archive.lco.global/}, 
we downloaded the data processed through the BANZAI pipeline\footnote{https://lco.global/documentation/data/BANZAIpipeline/}, which performs the standard data reduction of 
imaging data and aperture photometry using the SExtractor algorithm \citep{Bertin1996}. 
The data files contain flux values measured using circular apertures 
with several different fixed radii and those using an adaptively scaled 
elliptical aperture whose size is determined with the ``Kron'' radius 
\citep{kron1980}. To account for possible ellipticity of the point 
spread function (PSF) and dependency of PSF size and shape on the 
source flux and position on the image, we adopted the values 
from the adaptive elliptical aperture photometry 
and performed magnitude calibration using nearby stars, 
referring to the Pan-STARRS1 Surveys \citep{2016arX161205560C} 
for the g, r, i, and y bands and the second-generation 
Guide Star Catalog GSC 2.3 \citep{2008AJ....136..735L}
for the V and R bands.

\subsection{{\it Seimei}}

We performed optical spectroscopy using the 3.8-m {\it Seimei} telescope 
of Kyoto University at the Okayama Observatory \citep{krt10} 
once in the first rebrightening in 2019 \citep{yoshitake2022} 
and also at two nights in the 2020 rebrightening. In all these 
observations, we used the Kyoto Okayama Optical Low-dispersion 
Spectrograph with an integral field unit (KOOLS-IFU; \citep{ysd05,mtb19}), 
with the VPH-blue grism, whose wavelength coverage is 
4000--8900~\AA~and the wavelength resolution is 
$R=\lambda/\Delta \lambda \sim$ 500. 

The data reduction was conducted with IRAF and the pipeline 
tools specifically developed for the KOOLS-IFU\footnote{\url{http://www.kusastro.kyoto-u.ac.jp/~iwamuro/KOOLS/index.html}} downloaded 2023 July, which performs standard reduction processes including overscan and bias subtraction, flat fielding, wavelength calibration, spectral extraction, sky subtraction, and flux calibration. 
The Hg, Ne and Xe lamp data were used for the wavelength 
calibration. In the sky subtraction, we estimated the 
brightness of the sky from the same object frames using the 
data of fibers placed on a blank-sky area. The spectra 
obtained at the same night are averaged into one spectrum.

\begin{figure}[htbp]
  \begin{center}
  \includegraphics[width=80mm]{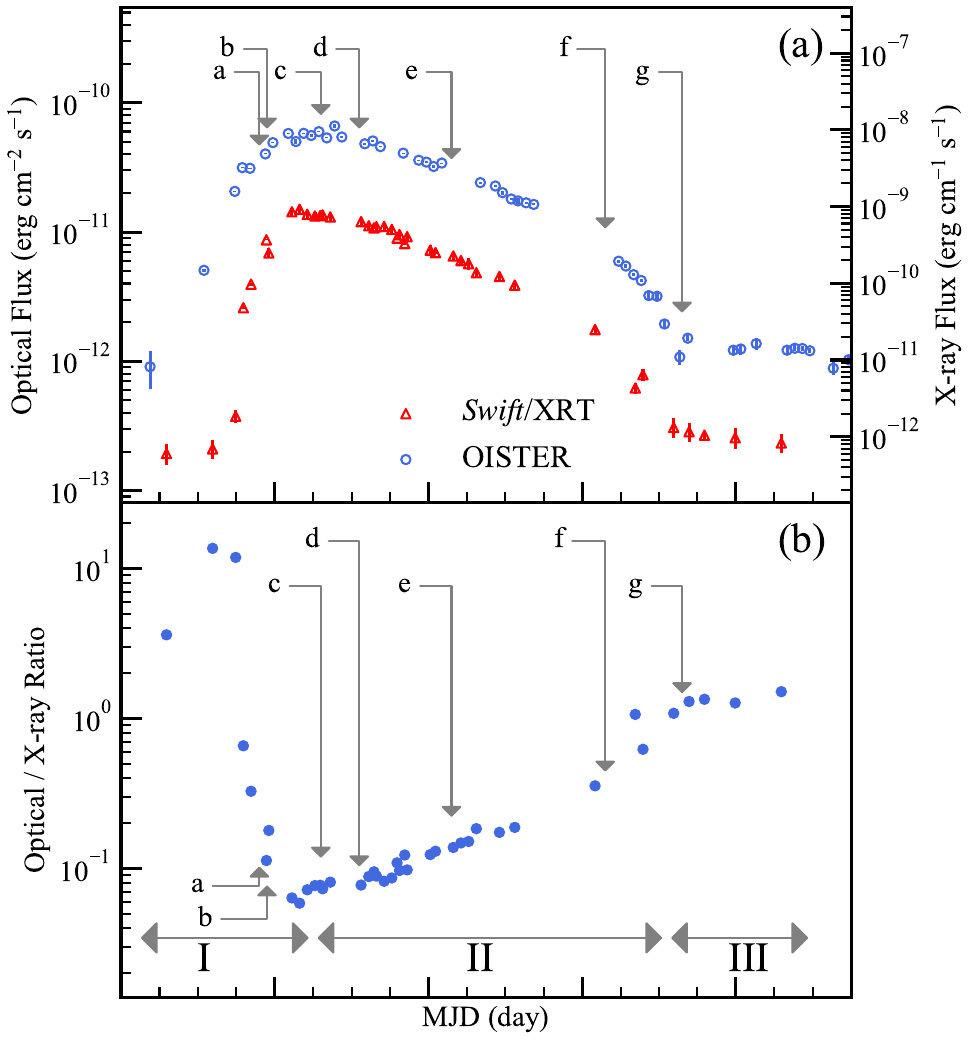}
  \end{center}
  \caption{(a) The g'-band light curve (blue circles) from the OISTER collaboration (Higuchi et al. in preparation) and X-ray light curve from the {\it Swift}/XRT (red triangles) in the first rebrightening. 
  The X-ray fluxes (in 0.4--10 keV) were estimated from the count rates  assuming a power-law shaped spectrum with a photon index of 1.6. The optical fluxes were obtained by multiplying the flux densities in Fig~\ref{fig:lc_opt} by the g'-band frequency ($6.2\times 10^{14}$ Hz). 
  (b) The optical versus flux ratio obtained from the data in the panel (a).  
  The horizontal arrows indicate the periods defined based on the behavior of the evolution of the fluxes and the ratio (Periods I--III; see text). 
  The alphabets a--g with broken arrows show the epochs at which the multi-wavelength SEDs in Figure~\ref{fig:unabsorbed_SED} are obtained.
  }
  \label{fig:lc_opt_x_ratio}
\end{figure}

\section{Analysis and Results}

\subsection{Long-term Light Curves}
In Figure~\ref{fig:lc_opt} we show long-term optical g'-band 
light curve for about 2 years after the discovery. 
After the main outburst in 2018, the source showed three rebrightenings. 
The X-ray data and the UV, optical, and near-IR photometric data were taken in the first (2019) rebrightening, while the {\it Seimei} observations were carried out in the same period and the third (2020) rebrightening. Note that the three rebrightening events have similar light curve profiles, as shown in Figure~\ref{fig:lc_opt}(b). 

Figure~\ref{fig:lc_opt_x_ratio} shows the enlarged view of the first rebrightening period, where X-ray light curve from the Swift/XRT is added. As seen in the top panel, the amplitude of the optical flux rise is larger than that of the X-ray flux. The bottom panel plots the 
evolution of the optical to X-ray flux ratio, which decreased 
rapidly during the rising phase, and then increased at a slower rate in the decaying phase. After a discontinuous jump, the flux ratio finally became almost constant in the last period of the rebrightening (where the optical flux declined somewhat more rapidly than before) and then back to the level before the rebrightening. 
We define these three periods with different behaviors in the optical versus X-ray ratio as Period I (before MJD 58565), Period II (MJD 58565--58611), and Period III (after MJD 58611), which correspond approximately to the flux rise, decay, and dim periods, respectively.

\subsection{Multi-wavelength SED}
\label{subsec:SEDana}

Using the X-ray spectra and the UV to near-IR photometric data listed in Table~\ref{tab:observation_x} and \ref{tab:observation_opt}, 
we created multi-wavelength SEDs on 2019 March 16, 17, 24, 29, 
Apr 9--10, Apr 30--May 1, and May 10--13 (hereafter we call 
Epoch a--g, in time order). 
We adopted the epochs that have X-ray data and at least 3 
near-IR/optical/UV band data. 
Epochs (a) and (b) are included in Period I and 
the remaining epochs are in Period II, 
except for Epoch (g), which is in Period III. 
Analysis of the SEDs was conducted with XSPEC 
version 12.13.0c, 
and the UV to near-IR data were converted to the XSPEC 
format using {\tt ftflx2xsp}. 

\begin{figure*}[htbp]
  \begin{center}
     \includegraphics[width=50mm]{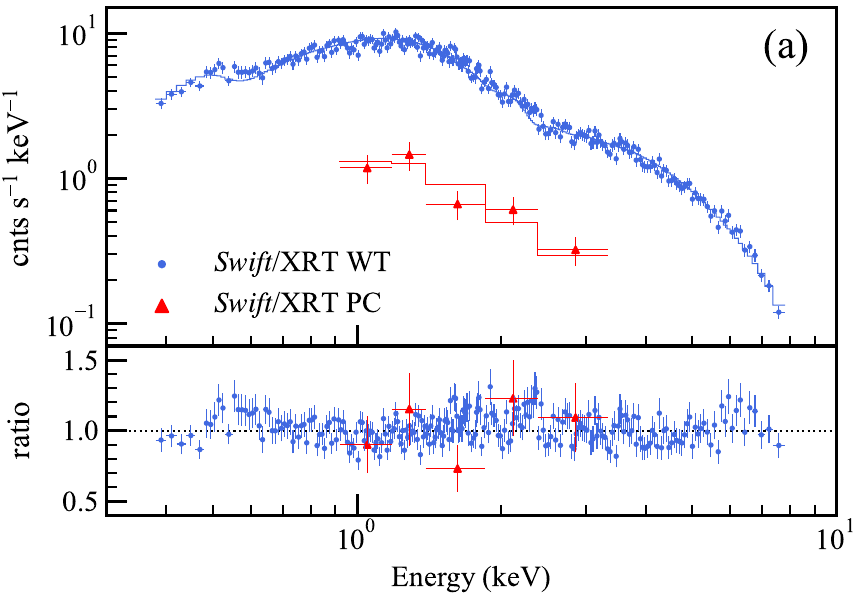}
      \includegraphics[width=50mm]{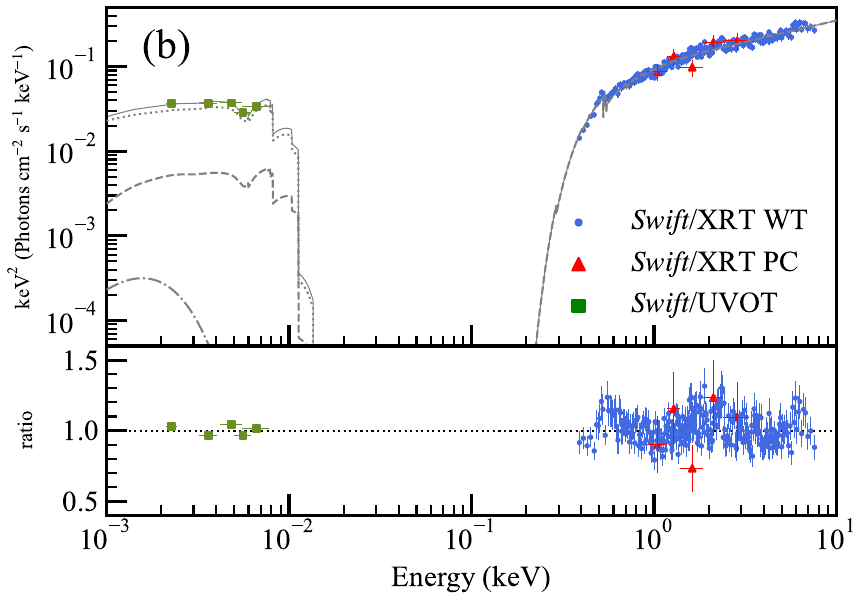}
    \includegraphics[width=50mm]{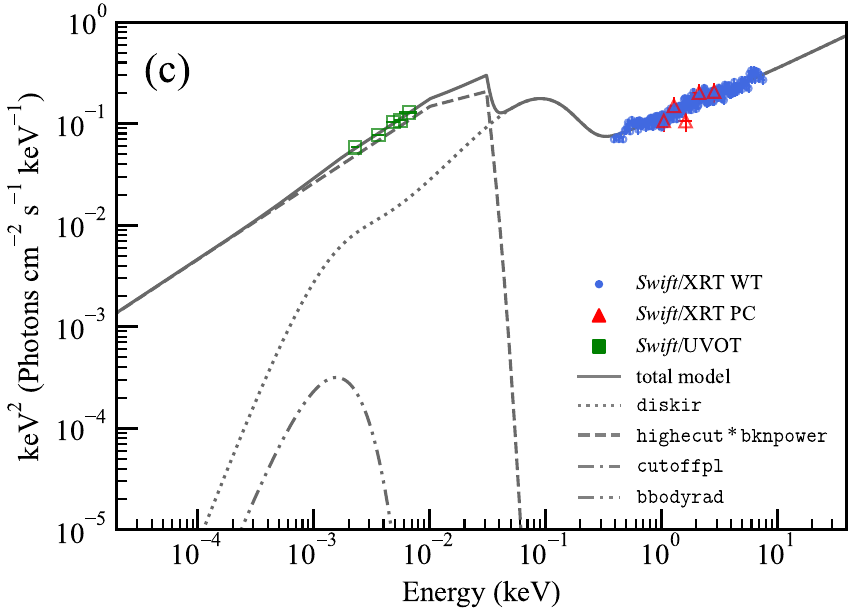}
  \end{center}
  \caption{ 
  (Left) X-ray spectrum in Epoch (c) and its best-fit absorbed power-law model. (Middle) multi-wavelength SEDs and best-fit {\tt highecut*bknpower+diskir+bbodyrad} model in Epoch (c). The bottom panel shows the data versus model ratio. (Right) same as the middle panel, but corrected for interstellar extinction. 
  Solid, dotted, dashed, and dot dashed lines show the total model and the {\tt diskir}, {\tt highecut*bknpower}, and {\tt bbodyrad} components, respectively. The blue circles, red triangles, green squares present the Swift/XRT data in the WT mode, those in the PC mode, and the Swift/UVOT data, respectively.}
  \label{fig:mar24fit} 
\end{figure*}

We first focused on the Epoch (c) data, which were obtained 
around the flux peak and have the best statistics and 
broadband coverage. 
As a first step, we considered the X-ray spectrum alone and 
fitted it with a simple absorbed power-law model. 
We adopted {\tt TBabs} for the absorption model, assuming the abundance table given by \citet{2000ApJ...542..914W}. As shown in Figure~\ref{fig:mar24fit}(a), this model was 
found to give an acceptable fit with a recuded chi-squared of 316 for 255 degrees of 
freedom. The best-fit photon index was $\Gamma = 1.53 \pm 0.03$, 
which is consistent with that in a typical LHS. 
We obtained the best-fit absorption column density $N_{\mathrm H}$ of 
($1.3 \pm 0.1) \times 10^{21}$ cm$^{-2}$, which is consistent with previously reported values using the Swift/XRT or NICER spectra at 
brighter states in the main outburst (e.g., \citealt{2018ATel11423....1U, sdt18, 2021ApJ...910...21R}) and rebrightenings \citep{sha21}. This value is also 
consistent with the Galactic column density estimated from 
the H I map ($N_\mathrm{H}=1.3 \times 10^{21}$ cm$^{-2}$, \citealt{2016AA...594A.116H}). 
In the following analysis, we fixed $N_{\mathrm H}$ of the {\tt TBabs} component at this value.

We then analyzed the multi-wavelength SED in Epoch (c). 
Here, we adopted the irradiated disk model {\tt diskir} \citep{gie08,gie09}, which is often used for multi-wavelength 
SEDs in the LHS, and actually employed 
in previous works of MAXI J1820 (e.g., \citealt{sdt18}; \citealt{ozbey2022}). 
The {\tt diskir} model considers the multi-color disk emission 
and its Comptonization, which illuminate the inner and outer disk regions. The irradiation strengths of the inner and outer regions are parameterized with $f_\mathrm{in}$ and $f_\mathrm{out}$, which is the fraction of bolometric flux which is thermalized in the inner and outer disk, respectively. Here, we ignored the illumination of the inner disk (i.e., $f_\mathrm{in}=0$). 
The $f_\mathrm{out}$ was allowed to vary within $1 \times 10^{-3}$  and $1 \times 10^{-2}$, a typical range of value in the LHS (e.g., \citealt{gie09}). 
The electron temperature of the Comptonization component was fixed at 65 keV, following \citet{ozbey2022}.  
We left the following parameters as free parameters: the inner disk temperature $\Tin$, the photon index $\Gamma$ of the Comptonization component, the ratio of luminosity $L_\mathrm{c}/L_\mathrm{d}$ in the Comptonization component to the unilluminated disk component, the ratio of outer and inner disk radius $\log(R_\mathrm{out}/R_\mathrm{in})$, and normalization. 
We set the upper limit of $R_\mathrm{out}$ to be 
the tidal truncation radius, $2.5 \times 10^6$ 
km\footnote{To realize this, we modified the {\tt diskir} code so that the normalization is parameterized by $R_{\rm out}$ instead of $R_{\rm in}$ (i.e.; ${\rm norm.} = (R_{\rm out} {\rm (km)})^2/D_{10} \cos i$, where $D_{10}$ represents the distance in units of 10 kpc.},  
which we assumed to be 0.8 times the Roche lobe size of 
the black hole (e.g., \citealt{2001LNP...573...69B}). 
The Roche lobe size can be estimated to be $3.2 \times 10^6$ km 
from the Kepler's third law and Equation (4) in \citet{pat71} 
assuming an orbital period of 0.68 day, a black hole mass of 
$7 M_\odot$ and a companion mass of $0.5 M_\odot$ \citep{tor20}. 

In addition, the {\tt bbodyrad} model is also incorporated to the model to 
account for the blackbody emission from the 
companion star. Here we assumed a blackbody temperature of 
$\sim 4700$ K and a radius of 0.65 $R_\odot$ considering the results from optical spectroscopy in a low luminosity state \citep{2019ApJ...882L..21T}.
We also added the {\tt TBabs} and {\tt redden} components to the full model to account for the interstellar absorption in the X-ray band and extinction in the near-IR to UV band, respectively. 
We assumed $E(B-V)$ for {\tt redden} to be $0.16$ \citep{2018ATel11418....1B}. 
Combining the above components, we obtained {\tt TBabs*redden*(diskir+bbodyrad)} model. 
Fitting the data with this model yielded a 
chi-squared value of 394 with 257 degree of freedom.

We also tested an alternative possibility: the UV-to-near-IR flux is dominated by the jet emission, instead of the irradiated outer disk emission as assumed in the {\tt diskir} model.
We adopted the broken power-law model ({\tt bknpower} in XSPEC) as a phenomenological model of the jet synchrotron spectrum, in which 
the power-law components above and below the break correspond to the optically thin and thick synchrotron emissions, respectively. Here, the photon index above the break was fixed at $\Gamma = 1.7$, which is a typical value of the optically thin synchrotron spectrum (e.g., \citealt{Russell2010}), while that below the break was allowed to vary. To account for the cooling break, we combined the multiplicative cutoff power-law model {\tt highecut}.
We assumed both the optically thin-to-thick spectral break and the cooling break of the synchrotron emission were located between UV and X-ray bands where no data are available, and fixed the break energy of {\tt bknpower} and the cutoff energy at $1\times 10^{-2}$ keV and $3\times 10^{-2}$ keV, respectively. 
We combined an additional power-law component to take into account the X-ray emission from hot inner accretion disk, and considered the interstellar absorption/extinction and blackbody emission from the companion star in the same manner as in the {\tt diskir} model. This model, {\tt TBabs*redden*(highecut*bknpower+powerlaw+bbodyrad)}, produced a much better fit than that of the {\tt diskir} model, with a chi-squared value of 329 for 256 degrees of freedom.
Figure~\ref{fig:mar24fit}(c) shows the data and the best-fit model and Table~\ref{tab:mar24fit} lists the best-fit parameters.

In the above analysis, we found that the observed SED was better described by the jet model than the irradiated disk model. In reality, however, the irradiated outer disk emission should be present as suggested by the optical emission lines detected in the rebrightening phases (see Section 3.3). We therefore tested also  
the combined jet plus irradiated disk model: 
{\tt TBabs*redden*(highecut*bknpower+diskir+bbodyrad)}. 
We found that the quality of fit obtained from this model ($\chi^2 = 329$ with degrees of freedom of 256) was not improved from the jet model. Table~\ref{tab:mar24fit} gives the best-fit parameters of this model. As shown in Figure~\ref{fig:mar24fit}(c), the contribution of the {\tt diskir} component to the optical and UV fluxes is a factor of $\gtrsim 5$ smaller than that of the jet synchrotron component. 

\begin{figure*}[htbp]
  \begin{center}
      \includegraphics[width=150mm]{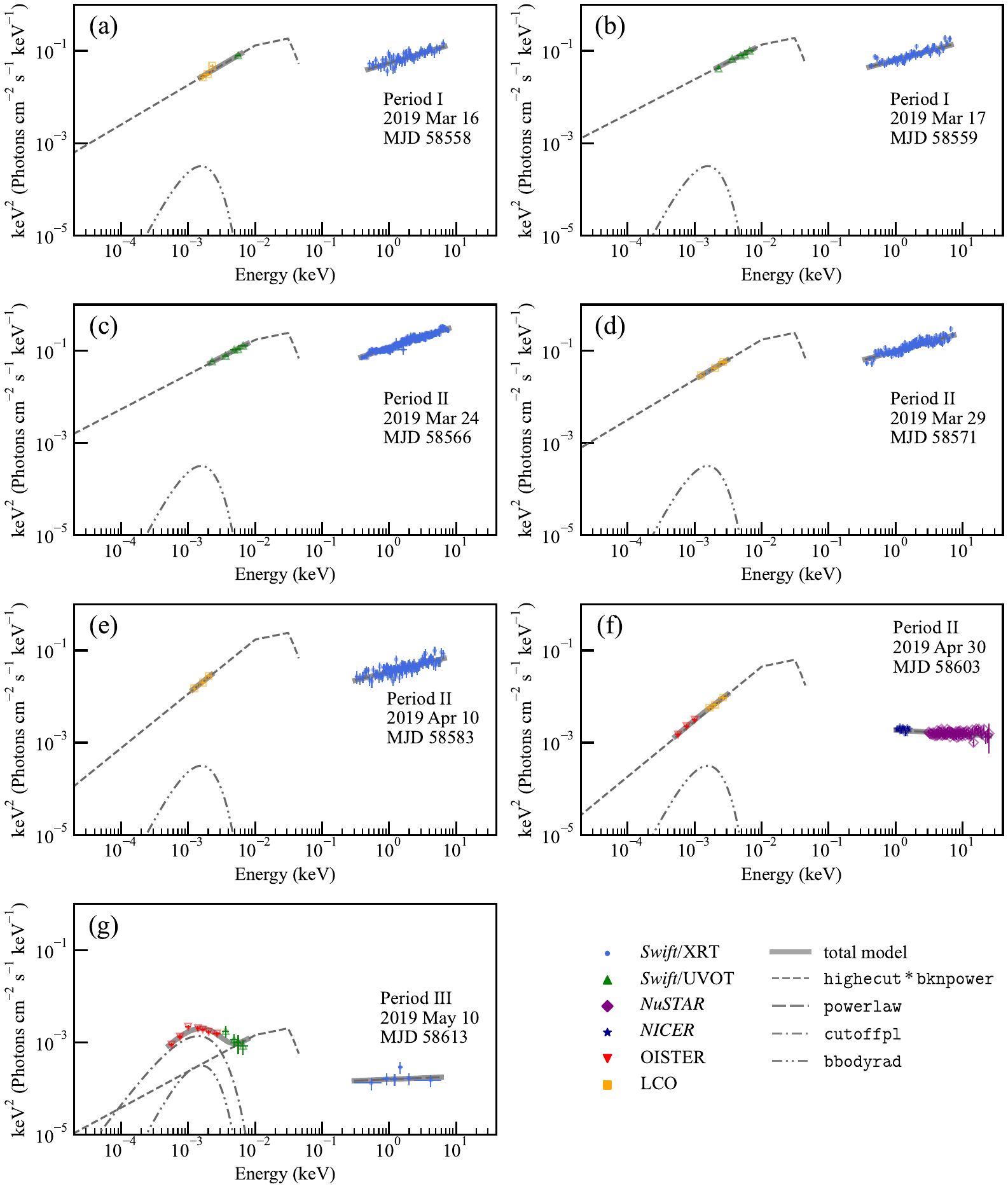} 
  \end{center}
  \caption{The multi-wavelength SEDs and best-fit models corrected for interstellar extinction. The {\it Swift}/XRT, {\it Swift}/UVOT, {\it NuSTAR}, {\it NICER}, OISTER, and LCO data 
  are shown in blue circles, green triangles, purple diamonds, dark blue stars, red inverse triangles, and orange squares, respectively. The adopted models are {\tt highecut*bknpower+powerlaw+bbodyrad} for Period I and II, and {\tt highecut*bknpower+cutoffpl+powerlaw+bbodyrad} for Period III. Solid,
  dashed, dot-double-dashed, dot dashed, and double-dot-dashed lines indicate the total model, and the 
  {\tt highecut*bknpower}, {\tt powerlaw}, {\tt cutoffpl}, and {\tt bbodyrad} components, respectively.}
  \label{fig:unabsorbed_SED}
\end{figure*}

Next, we performed SED fitting 
for the rest of the data in Period I and II, using 
the jet model ({\tt TBabs*redden*(highecut*bknpower+powerlaw+bbodyrad)}). We note that the {\tt diskir} plus jet model was also tested but the fit was not improved significantly and resulted in a negligible contribution of the {\tt diskir} component to the optical and UV bands. 
We fixed $N_\mathrm{H}$ of {\tt TBabs} at $1.3 \times 10^{21}$ cm$^{-2}$ 
as we did for the SED in Epoch (c).
As shown in Figure~\ref{fig:unabsorbed_SED}, the model successfully reproduced 
all the Period-I and Period-II data. 
The best-fit parameters are listed in Table~\ref{tab:SEDpars}. 

The Epoch (g) were analyzed in \citet{yoshitake2022},
where we found that the SED profile is best interpreted by 
the ADAF model.  
Following \citet{yoshitake2022} 
we used the cutoff power-law model ({\tt cutoffpl}) as an 
approximated model of the synchro-cyclotron emission from 
the ADAF, which mainly contributes to the optical band, 
and a power-law model to represent the thermal bremsstrahlung and/or 
Comptonization of the synchro-cyclotron emission. 
Adding a jet component and the blackbody emission from the 
companion star in the same way as Period-I and II data, 
we adopted the 
{\tt TBabs*redden(highecut*bknpower+cutoffpl+powerlaw+\\bbodyrad)} 
model. Figure~\ref{fig:unabsorbed_SED}(g) presents the 
SED data and the best-fit model and Table~\ref{tab:SEDpars} 
lists the best-fit parameters, which are 
consistent with those in \citet{yoshitake2022}. 

\begin{table*}
    \footnotesize
\begin{threeparttable}
    \centering
    \tbl{Best-fit parameters of the SED fitting in Epoch (c), with the jet and {\tt diskir} model}{
    \begin{tabularx}{0.85\linewidth}{cc cc cc}
    \hline \hline
\multirow{8}{*}{Epoch}
& \multirow{1}{*}{Date}
& $N_\mathrm{H}$ 
& Flux$_\mathrm{opt}^\mathrm{jet}$\tnote{*}
& $\Gamma_\mathrm{jet}$
& $L_\mathrm{opt}^\mathrm{jet}$\tnote{*}
\\
&
& $10^{21}$ cm$^{-2}$
& $10^{-12}$ erg cm$^{-2}$ s$^{-1}$
&
& erg s$^{-1}$
\\ \\
& \multirow{1}{*}{MJD}
& $T_\mathrm{in}$ 
& $f_\mathrm{out}$
& $\Gamma_\mathrm{diskir}$
& $L_\mathrm{X}$\tnote{*}
\\
&
& keV
&
&
& erg s$^{-1}$
\\ \\
& $\chi^2\mathrm{/d.o.f}$
& $R_\mathrm{out}$  
& $R_\mathrm{out} / R_\mathrm{in}$
& $R_\mathrm{in}$ 
& $L_\mathrm{c}/L_\mathrm{d}$
\\
&
& $r_\mathrm{g}$\tnote{$\dagger$}
&
& $r_\mathrm{g}$
&
\\ \hline
\multirow{3}{*}{(c)}
& \multirow{1}{*}{2019 Mar 24}
& $ 1.3 $ (fixed)
& $2.1_{-0.2}^{+0.1} \times 10^{2}$
& $1.3\pm0.1$ 
& $1.6\times 10^{35}$
\\
& 58566
& $ \left(3.1\pm0.1\right)  \times 10^{-2}$
& $1.0^{+2.5}_{-0\mathrm{(pegged)}} \times 10^{-3}$
& $ 1.54\pm 0.02 $
& $6.0\times 10^{35}$
\\
& $328 / 256$
& $ 2.4_{-1.1}^{+0\mathrm{(pegged)}} \times 10^{5}$
& $1.0_{-0 \mathrm{(pegged)}}^{+1.0} \times 10^{3}$
& $2.4_{-1.7}^{+0 \mathrm{(pegged)}} \times 10^2$
& $ 9.9_{-5.6}^{+0.1 \mathrm{(pegged)}} $
    \\ \hline
    \end{tabularx}}
    \smallskip
    \begin{tablenotes}[normal]
    \footnotesize
    \item {\tt TBabs*redden*(highecut*bknpower+diskir+bbodyrad)}, 
      with $N_{\mathrm H}$ of {\tt TBabs} and $E(B-V)$ of {\tt redden} fixed at $1.3 \times 10^{21}$ cm$^{-2}$ and $0.16$, respectively. For the {\tt bbodyrad} component,
      $T_{\rm bb} = 4700$ K and $R_{\rm bb} = 0.65 R_\odot$, and $D = 3$ kpc were assumed.
      For the {\tt diskir} component, the electron temperature of the Comptonization 
      component was fixed at 65 keV, 
      the fraction of bolometric flux thermalized in the inner disk $f_\mathrm{in}$ at $1.2 \times 10^{-3}$,        
      and the radius of the Compton illuminated disk in terms of the inner disk radius at $1.005$.
    We assume $D = 3$ kpc, $i = 70^\circ$, and $M = 7 M_\odot$ in order to estimate $R_\mathrm{in}$, $R_\mathrm{out}$, and the luminosity of the {\tt diskir} component.
      \item[*]%
      Unabsorbed luminosity and flux of each component in 0.5--5 keV and 0.5--5 eV. 
    \item[$\dagger$] %
      Gravitational radius $r_\mathrm{g} \equiv GM/c^2$ (where $G$ is the gravitational constant and $c$ is the light velocity). 
      $r_\mathrm{g}$ is $ 10.5$ km assuming $M=7 M_\odot$.
    \end{tablenotes}
    \label{tab:mar24fit}
  \end{threeparttable}
\end{table*}

\begin{table*}
    \footnotesize
\begin{threeparttable}
    \centering
    \tbl{Best-fit parameters of the SED fitting with the jet model}{
    \begin{tabularx}{0.9\linewidth}{cc cc cc}
    \hline \hline
\multirow{8}{*}{Epoch}
& \multirow{1}{*}{Date}
& $N_\mathrm{H}$ 
& $\Gamma_\mathrm{jet}$
& Flux$_\mathrm{opt}^\mathrm{jet}$\tnote{*}
& $L_\mathrm{opt}^\mathrm{jet}$\tnote{*} 
\\
&
& $10^{21}$ cm$^{-2}$
&
& $10^{-12}$ erg cm$^{-2}$ s$^{-1}$
& erg s$^{-1}$
\\ \\
& \multirow{1}{*}{MJD}
& 
& $\Gamma_\mathrm{brems}$
& norm$_\mathrm{brems}$
& $L_\mathrm{X}$\tnote{*} 
\\
&
& 
& 
&
& erg s$^{-1}$
\\ \\
& $\chi^2\mathrm{/d.o.f}$
& 
& $E_\mathrm{peak, ADAF}$
& norm$_\mathrm{ADAF}$
& 
\\
&
& 
& keV
& 
& 
\\ \hline
\multicolumn{6}{c}{Period I (rise)} \\
\multirow{3}{*}{(a)}
& \multirow{1}{*}{2019 Mar 16}
& $ 1.3 $ (fixed)
& $1.1\pm0.1$
& $\left(1.3\pm0.1\right)\times 10^{2}$
& $1.3\times 10^{35}$
\\
& 58558
& 
& $1.6\pm 0.1$
& $ 2.0_{-0.5}^{+0.6} \times 10^{15} $
& $2.8\times 10^{35}$
\\
& $42/49$
& 
& -
& -
& 
\\
\multirow{3}{*}{(b)}
& \multirow{1}{*}{2019 Mar 17}
& $ 1.3 $ (fixed)
& $1.2\pm0.1$
& $\left(1.6\pm0.1\right)\times 10^{2}$
& $1.5\times 10^{35}$
\\
& 58559
& 
& $1.6\pm 0.1$
& $ 2.6_{-0.4}^{+0.5} \times 10^{15} $
& $3.1\times 10^{35}$
\\
& $68/62$
& 
& -
& -
& 
\\
\multicolumn{6}{c}{Period II (decay)} \\
\multirow{3}{*}{(c)}
& \multirow{1}{*}{2019 Mar 24}
& $ 1.3 $ (fixed)
& $1.2\pm0.1$
& $\left(2.0\pm0.1\right)\times 10^{2}$
& $1.9\times 10^{35}$
\\
& 58566
& 
& $1.53\pm 0.02$
& $ \left(3.9\pm 0.2\right)\times 10^{15} $
& $6.0\times 10^{35}$
\\
& $327/259$
& 
& -
& -
& 
\\
\multirow{3}{*}{(d)}
& \multirow{1}{*}{2019 Mar 29}
& $ 1.3 $ (fixed)
& $1.1\pm0.4$
& $\left(1.7\pm0.2\right)\times 10^{2}$
& $1.6\times 10^{35}$
\\
& 58571
& 
& $1.60\pm 0.03$
& $ \left(3.9\pm0.4\right)\times 10^{15} $
& $4.8\times 10^{35}$
\\
& $152/128$
& 
& -
& -
& 
\\
\multirow{3}{*}{(e)}
&\multirow{1}{*}{2019 Apr 10}
& $ 1.3 $ (fixed)
& $0.8_{-0.6}^{+0.5}$
& $1.1_{-0.3}^{+0.4} \times 10^{2}$
& $1.0\times 10^{35}$
\\
& 58583
& 
& $1.6\pm 0.1$
& $ \left(1.5\pm 0.3 \right)\times 10^{15} $
& $1.6\times 10^{35}$
\\
& $79/102$
& 
& -
& -
& 
\\
\multirow{3}{*}{(f)}
&\multirow{1}{*}{2019 Apr 30}
& $ 1.3 $ (fixed)
& $0.81\pm0.13$
& $\left(2.7\pm0.3\right)\times 10^{1}$
& $2.6\times 10^{34}$
\\
& 58603
& 
& $2.1\pm 0.03$
& $ \left(2.3 \pm 0.3 \right)\times 10^{4} $
& $7.2\times 10^{33}$
\\
& $94/87$
& 
& -
& -
& 
\\
\multicolumn{6}{c}{Period III (dim)}\\
\multirow{3}{*}{(g)}
&\multirow{1}{*}{2019 May 10--13}
& $ 1.3 $ (fixed)
& $1.2_{-2.0}^{+0.6}$
& $1.5_{-1.3}^{+1.4}$
& $5.2\times 10^{33}$
\\
& 58613
& 
& $1.9\pm 0.3$
& $ 3.5_{-1.7}^{+2.9} \times 10^{4} $
& $6.5\times 10^{32}$
\\
& $13/11$
& 
& $7.1_{-0.8}^{+1.5}\times 10^{-4}$
& $4.8_{-2.4}^{+1.5}\times 10^{4}$
& 
\\
    \\ \hline
    \end{tabularx}}
    \smallskip
    \begin{tablenotes}[normal]
    \footnotesize
    \item {\tt TBabs*redden*(highecut*bknpower+powerlaw+bbodyrad)} (Period I and II) and
      {\tt TBabs*redden*(highecut*bknpower +cutoffpl+bbodyrad)} (Period III). The save values as those in Table~\ref{tab:mar24fit} were adopted for the fixed parameters (see text).   
      \item[*]%
      Unabsorbed luminosity and flux of each component in 0.5--5 keV  and 0.5--5 eV. 
    \end{tablenotes}
    \label{tab:SEDpars}
  \end{threeparttable}
\end{table*}

\subsection{Optical Spectrum}
\label{subsec:seimei_ana}

\begin{figure*}[htbp]
  \begin{center}
     \includegraphics[width=150mm]{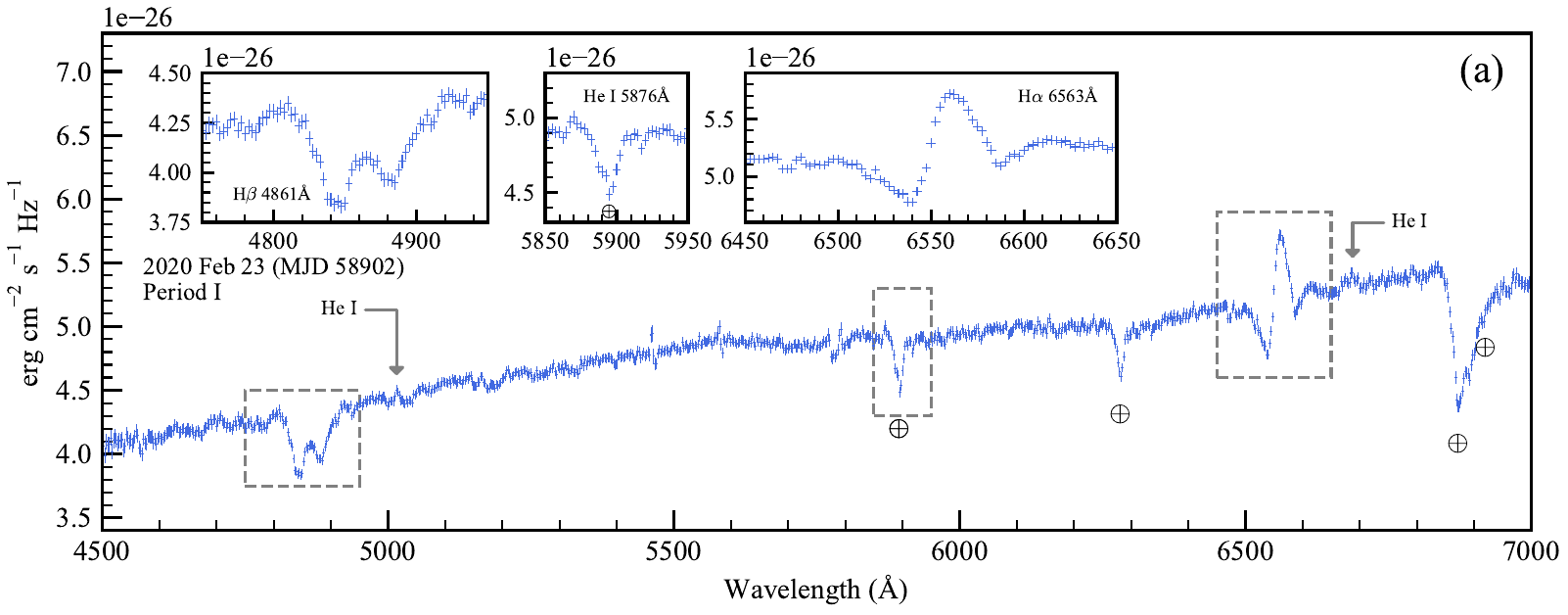}
     \includegraphics[width=150mm]{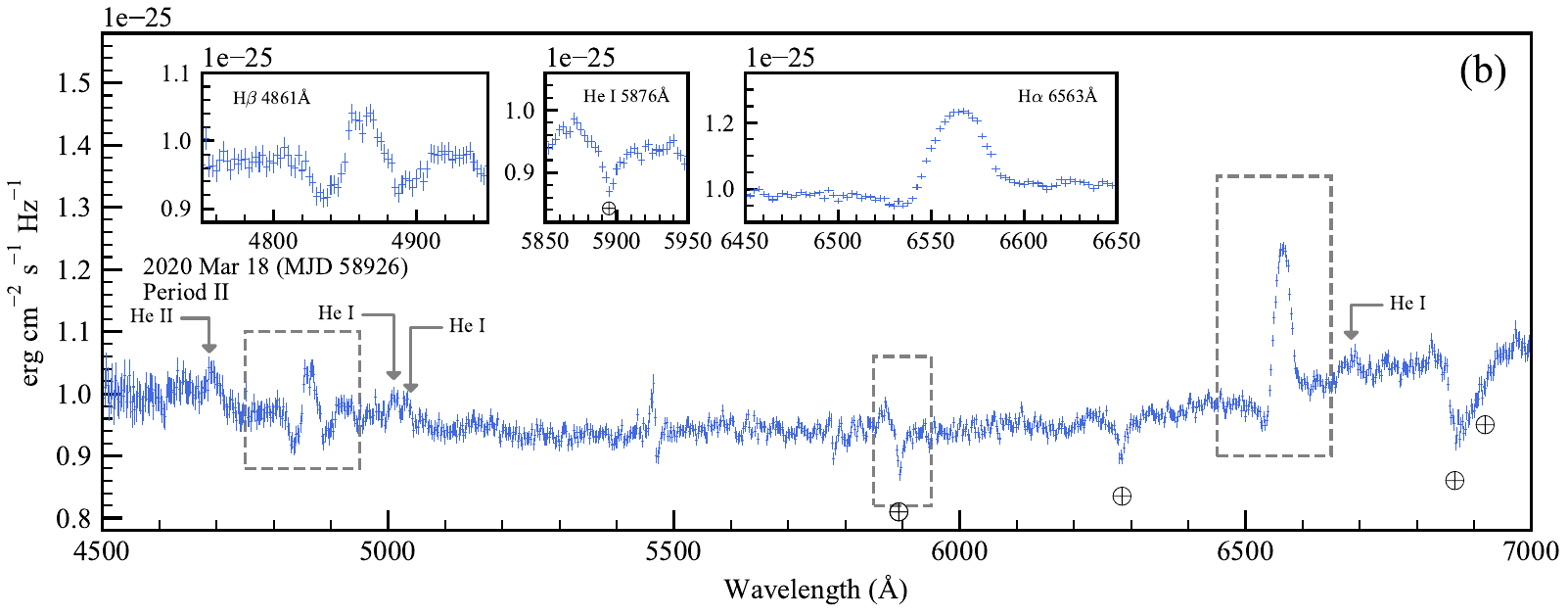}
     \includegraphics[width=150mm]{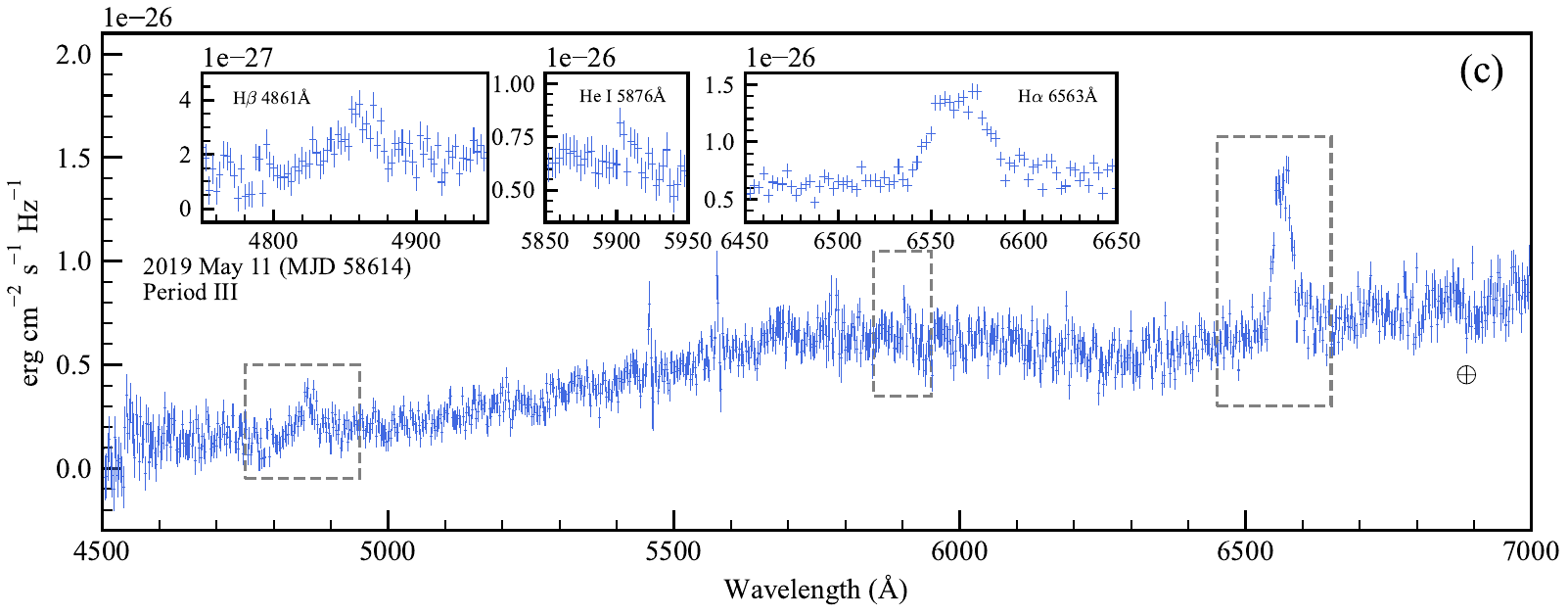}
  \end{center}
  \caption{{\it Seimei} spectra obtained in the three epochs. The panels (a), (b), and (c) correspond to Period I, II, and III, respectively. The inset panels show enlarged views around H$\alpha$ ($\lambda=6563$~\AA), H$\beta$ ($\lambda=4861$~\AA), 
  and He I ($\lambda=5876$~\AA) lines. Other identified lines are also shown with grey arrows.
  }
  \label{fig:seimei_fullSpectrum}
\end{figure*}

In Figure~\ref{fig:seimei_fullSpectrum}  
we show the {\it Seimei} spectra obtained in the three epochs, in which the H$\alpha$ ($\lambda=6563$~\AA) and H$\beta$ ($\lambda=4861$~\AA) lines are clearly detected. A weak He I ($\lambda=5876$~\AA) emission line is also seen on 2020 Feb 23 and March 18, and other He I and He II emission lines are identified especially on 2020 Mar 18. These He lines are weaker than the Balmer lines, and the He I line at $\lambda=5876$~\AA is partly overlapped with the atmospheric Na D absorption features ($\lambda=5890$~\AA~and $\lambda=5896$~\AA), making it difficult to separate the individual contributions and determine the He I line profile accurately. We therefore focused only on the two Balmer lines and investigated their spectral profiles.

\begin{figure*}[htbp]
  \begin{center}
     \includegraphics[width=150mm]{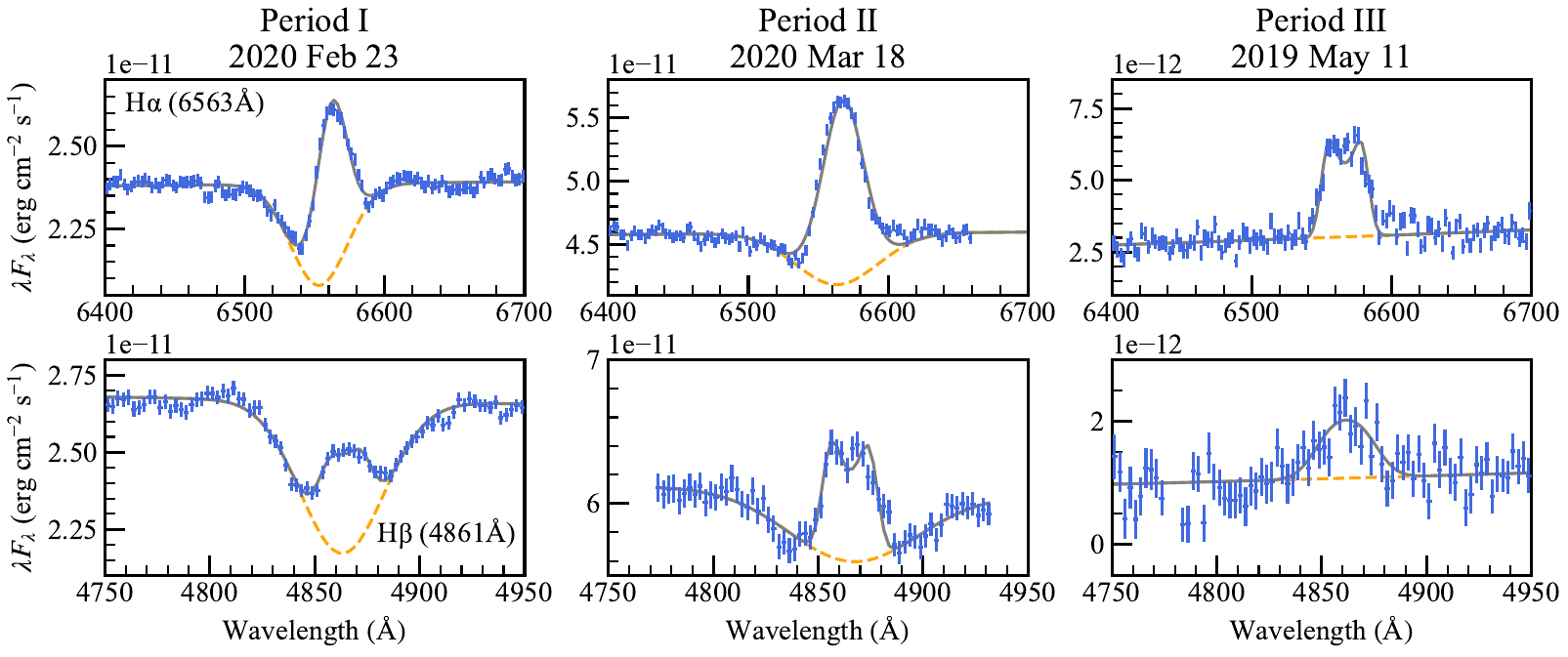} 
  \end{center}
  \caption{{\it Seimei}/KOOLS-IFU spectra around the H$\alpha$ and H$\beta$ lines and their best-fit models. Gray solid and orange dashed lines present the total model and the sum of the continuum and absorption line components, respectively. Note that the data and the model include broadening due to the response of the grism.}
  \label{fig:seimei_optSED}
\end{figure*}

Figure~\ref{fig:seimei_optSED} shows the H$\alpha$ and 
H$\beta$ spectra obtained with the {\it Seimei} telescope. 
On 2019 May 11, which is 47 days after 
the g'-band flux peak and corresponds to Period III, 
only emission lines are visible, whereas 
broad absorption structures are seen in H$\alpha$ 
and H$\beta$ spectra on 2020 Feb 23. On 2020 March 18 
the absorption structures still exist but less 
significant than those in Feb 23. 
Considering the similarity of the optical g'-band light 
curve in 2019 and 2020 rebrightening events, we 
investigated which of three periods (Period I, II, and III) 
the two observations in 2020 correspond to. 
On the basis of the number of days from the g'-band flux peak, 
we found that the observations on 2020 Feb 23 and March 18
are $-7$ days and $+17$ days from the peak and can be included in Period I and II, respectively (see Figure~\ref{fig:lc_opt}b).  

To characterize the line profiles, we performed model 
fitting of the H$\alpha$ and H$\beta$ spectra on XSPEC. 
We adopted a wavelength range of 6400--6700~\AA~for the 
H$\alpha$ data and 4750--4950~\AA~for the 
H$\beta$ data and converted them with the ftool 
{\tt ftflx2xsp} to the XSPEC format. The response matrix 
files were generated with the ftool {\tt ftgenrsp}, 
which assumes a Gaussian response function for each 
wavelength bin. 
To account for the spectral resolution of the 
{\it Seimei}/KOOLS-IFU, we adopted a full width at half 
maximum (FWHM) of 9.41~\AA, 8.29~\AA, and 11.14~\AA~in the H$\alpha$ spectra
and 19.24~\AA, 12.61~\AA, and 7.35~\AA~in the H$\beta$ spectra
for the Gaussian response functions on 2019 May 11, 2020 Feb 23,
and 2020 Mar 18, respectively. These values were 
estimated from the profile of a Ne line around the 
H$\alpha$ line and a Hg 
line around H$\beta$ line in the arc lamp frame.

We fitted the continuum with a single power-law model 
and adopted a Gaussian model with a negative normalization for the broad 
absorption component. Some of the emission lines, 
especially the H$\alpha$ line on 2019 May 11 and 
the H$\beta$ lines, show a clear flat-topped or 
double-peaked profile, which is often seen in 
the emission lines from the accretion disk. 
The peak-to-peak separation of the H$\beta$ 
line on 2020 March 18, which most clearly shows a 
double-peaked profile, was estimated to be 
$17\pm 3$~\AA.\noprint{**} For these emission line 
components, we employed the {\tt diskline} model 
\citep{fabian1989}. 
The {\tt diskline} model calculates the emission line 
profile originating from the Keplerian accretion disk 
illuminated by the X-rays from the inner disk region. 
We note that the {\tt diskline} model calculate 
the line profile in almost the same way as the \citet{horne1986} model, often used in the optical line modelling; it computes the Doppler shifts of the individual part of the accretion disk and integrates their contributions  over the entire line emitting region of the disk, assuming that the emissivity depends on the radius $r$ as $r^{\alpha}$. 
The {\tt diskline} model considers the relativistic 
effects in the vicinity of the black hole, but they 
are negligible at large radii.
Following \citet{yoshitake2022}, we adopted 
$i = 70^{\circ}$ and $\alpha = -3$.
The outer disk radius $R_\mathrm{out}$ of the emitting 
region is assumed to be $2.4 \times 10^5 r_{\rm g}$, 
which is estimated in 
Section~\ref{subsec:SEDana}. We allowed the inner 
disk radius $R_\mathrm{in}$ to vary. 
The H$\alpha$ lines on 2020 Feb 23 and Mar 18 
were fitted with a Gaussian model. 

Figure~\ref{fig:seimei_optSED} shows the results 
of the data and their best-fit models.
Table~\ref{tab:SEDseimei} gives the line widths 
(half width at zero intensity; HWZI), the 
equivalent widths, and the inner 
radius $R_\mathrm{in}$ of the line-emitting/absorbing 
regions estimated in the fit. For Gaussian components 
we adopted their 5$\sigma$ values as the HWZIs and 
calculated $R_\mathrm{in}$ assuming the Kepler motion as  
\begin{equation}
R_\mathrm{in} \approx 0.87 \left(\frac{\sin i}{\sin 70^{\circ} }\right)^2 
\left(\frac{\mathrm{HWZI}}{\lambda_\mathrm{rest}}\right)^{-2} r_g, 
\label{eq:kepler}
\end{equation} where $\lambda_\mathrm{rest}$ is 
the rest frame wavelength of the line. 
For the {\tt diskline} component, we adopted the 
best-fit $R_\mathrm{in}$ value and estimated the 
HWZI value using Equation~\ref{eq:kepler}. 

As shown in Figure~\ref{fig:seimei_optSED}, the line centers of the H$\alpha$ absorption line in Epoch II and the H$\beta$ absorption lines in Epoch I and II are consistent with the rest frame wavelengths. The exception is the H$\alpha$ line on 2020 Feb. 23 (Period I), which shows a blueshift. However, the shift is only within $\sim$ 10 \AA~(corresponding to $\lesssim$ 500 km s$^{-1}$), which is somewhat smaller than the typical value ($\gtrsim$ 1000 km s$^{-1}$) of disk winds detected in the optical band (e.g., \citealt{MunozDarias2016}), and the absorption line is deep and wide compared with usual P Cygni profiles. 
As seen in Figure~\ref{fig:seimei_fullSpectrum}, the He I line show no significant blueshifted absorption component, either.

\begin{table*}
\begin{threeparttable}
    \centering
    \tbl{The best-fit line widths and equivalent widths of H$\alpha$ and H$\beta$ lines, and inner radii of the their emitting/absorbing regions}{
    \begin{tabularx}{\linewidth}{cc c cc cc c}
    \hline \hline
\multirow{2}{*}{Period} 
& \multirow{1}{*}{Date}
& \multicolumn{2}{c}{\multirow{2}{*}{Line}}
& \multirow{2}{*}{HWZI (\AA)}
& \multirow{2}{*}{Equivalent Width (\AA)}
& \multirow{2}{*}{$R_\mathrm{in}$ ($r_g$)}
\\
& Time Relative To Peak (day)
& 
&  
& 
\\ \hline
\multirow{4}{*}{I} 
& \multirow{2}{*}{2020 Feb 23}
& \multirow{2}{*}{H$\alpha$}
& Emis. 
& $ 53 \pm 1 $	
& $6.1\pm 0.2$ 
& $ \left( 1.3 \pm 0.1 \right) \times 10^{4} $\\
&
& 
& Abs.
& $ 99_{-3}^{+5} $	
& $6.5 \pm 0.2$
& $  3.8_{-0.4}^{+0.2}  \times 10^{3} $	
\\
& \multirow{2}{*}{$-7$}
& \multirow{2}{*}{H$\beta$}
& Emis.
& $ 11_{-2}^{+1} $
& $3.2 \pm 0.7$
& $ 1.7_{-0.3}^{+0.8}  \times 10^{5} $ \tnote{$\dagger$}
\\
&
&
& Abs.
& $ 96_{-6}^{+7} $	
& $9.2 \pm 0.7$
& $ \left( 2.3 \pm 0.3 \right) \times 10^{3} $	
\\
\multirow{4}{*}{II} 
& \multirow{2}{*}{2020 Mar 18}
& \multirow{2}{*}{H$\alpha$}
& Emis. 
& $ 66 \pm 1 $	
& $12.3 \pm 0.2$
& $  8.7_{-0.4}^{+0.3}  \times 10^{3} $	
\\
&
&
& Abs.
& $ \left( 1.1 \pm 0.1 \right) \times 10^{2} $	
& $6.5\pm0.4$
& $  2.9_{-2}^{+1}  \times 10^{3} $
\\
& \multirow{2}{*}{$+17$}
& \multirow{2}{*}{H$\beta$}
& Emis.
& $ 14_{-1}^{+2} $
& $3.5 \pm 0.3$ 
& $  1.1_{-0.2}^{+0.1}  \times 10^{5} $	
\tnote{$\dagger$}
\\
&
&
& Abs.
& $ \left( 1.5 \pm 0.1 \right) \times 10^{2} $	
& $6.2\pm 0.8$
& $  8.8_{-1}^{+1}  \times 10^{2} $	
\\
\multirow{4}{*}{III} 
& \multirow{2}{*}{2019 May 11}
& \multirow{2}{*}{H$\alpha$}
& Emis. 
& $ 18 \pm 1 $	
& $36 \pm 2$
& $  1.2_{-0.1}^{+0.2}  \times 10^{5} $	
\tnote{$\dagger$}
\\
&
&
& Abs.
& - & - & -
\\
& \multirow{2}{*}{$+47$}
& \multirow{2}{*}{H$\beta$}
& Emis.
& $ 13_{-5}^{+7} $	
& $26 \pm 5$
& $  1.2_{-0.7}^{+1.8\mathrm{(pegged)}}  \times 10^{5} $
\\
&
&
& Abs.
& - & - & -
\\
    \hline
    \end{tabularx}}
    \smallskip
     \begin{tablenotes}[normal]
      \footnotesize
      \small
    \item[$\dagger$] %
      The inner radii $R_\mathrm{in}$ with the dagger marks are estimated with {\tt diskline}, while the others are calculated from 5$\sigma$ width of the Gaussian model assuming a Keplerian rotation (see text).  
    \end{tablenotes}
    \label{tab:SEDseimei}
\end{threeparttable}
\end{table*}

\begin{figure*}[htbp]
  \begin{center}
     \includegraphics[width=80mm]{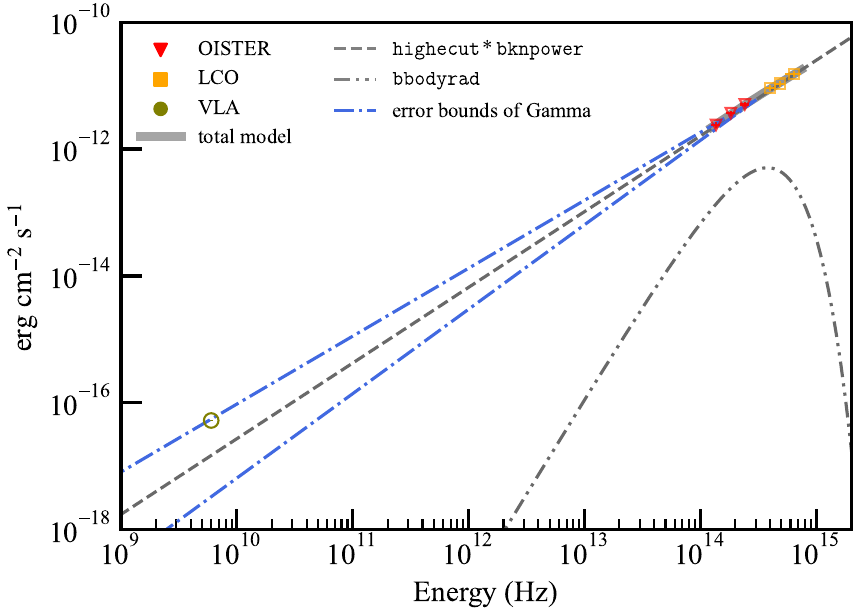} 
  \end{center}
  \caption{Optical-UV SED and the best-fit model in Epoch (f) (same as Figure~\ref{fig:unabsorbed_SED}(f) but the X-ray band is ignored), with radio data obtained with VLA on the same day. The black solid line shows the best-fit broken power-law component (below the break) for the jet, which is extrapolated to the radio band. The blue dash dotted lines are obtained by changing the photon index to its 90\% upper and lower limits.}
  \label{fig:unabsorbed_SED_radio}
\end{figure*}

\begin{figure}[htbp]
  \begin{center}
     \includegraphics[width=80mm]{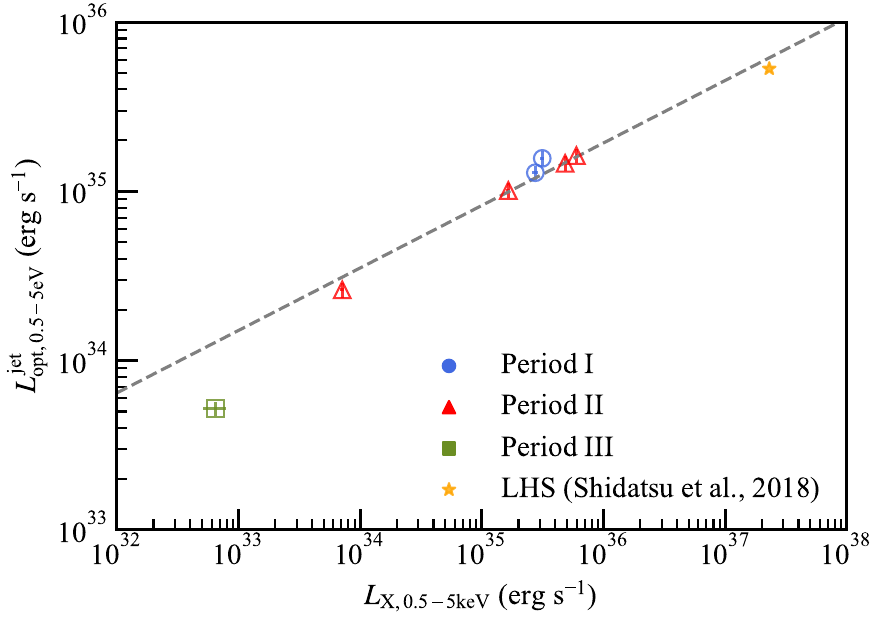} 
  \end{center}
  \caption{Correlation of the unabsorbed X-ray luminosity and optical jet luminosity of MAXI J1820. The blue circles, red triangles, green squares, and yellow star show the data taken in Period I, II, and III, and in the LHS of the main outburst \citep{sdt18}, respectively. }
  \label{fig:lumin_ratio}
\end{figure}

  \section{Discussion}
\label{sec:discussion}

\subsection{SED}

\subsubsection{Periods I and II}

We find that all X-ray spectra in Periods I and II are well
represented with a power law model modified with interstellar
absorption, confirming previous works on a part of the data 
analyzed here \citep{2021ApJ...907...34S, yoshitake2022, ozbey2022}.
The power-law dominant X-ray spectra indicate that MAXI J1820 
was always in the LHS and 
no transition to the HSS took place, 
which is also evident from the hardness ratio estimated in 
Figure~1 of \citet{stiele_and_kong2020}.
The LHS is consistent with the low Eddington ratios throughout the
rebrightening phase, $\log \lambda_{\rm Edd}< -3$\noprint{xxx}; 
usually transition from the LHS to the HSS in the initial outburst 
phase of a BHXB takes place typically at 
$\log \lambda_{\rm Edd} > -2$\noprint{xxx} 
(e.g., \citealt{2003AA...409..697M, 2019MNRAS.485.2744V}).
The photon index was about 1.6 
in Epochs (a)--(e) when the source was relatively bright 
($\log L_{\rm X}/L_{\rm Edd} \sim -3$\noprint{xxx}), whereas it
became larger (2.1) at a fainter state 
($\log L_{\rm X}/L_{\rm Edd} 
\sim -6$\noprint{xxx}). This trend agrees with past studies 
reporting negative correlations between the photon index 
and Eddington ratio at $-6.5 <
\log L_{\rm X}/L_{\rm Edd} <-3$ in BHXBs \citep{2015MNRAS.447.1692Y}.

We have shown that the near-IR/optical/UV SEDs in Periods I
and II are reproduced by the sum of 
the blackbody component from the companion star
and a dominant 
power-law component with a partial contribution
from the irradiated disk.
The photon indices of the power-law components
in Epochs (a)--(d), $\Gamma=$1.1--1.4
(corresponding to the energy indices
$\alpha=$0.1--0.4\footnote{$F_\nu \propto \nu^{-\alpha}$ where $F_\nu$ is the energy flux at frequency $\nu$}),
are consistent with that of the averaged SED during
the initial outburst reported by \cite{2021MNRAS.504.3862T}.
In Epochs (e) and (f), the photon indices are somewhat smaller,
$\sim$0.7--0.8.
The reason is unclear; synchro-cyclotron emission from
the hot flow,
similar to that predicted by an ADAF model (see
Section~\ref{subsec:Disc_SEDPeriod3}), might partially contribute to
the optical flux at high frequencies in these epochs.

We interpret that the origin of the optical power-law component 
is synchrotron radiation from jets. In fact, 
the energy indices we have obtained, $-0.4 \lesssim \alpha \lesssim -0.3$, are
consistent with that expected from an optically thick jet
($\alpha \sim 0$ for a Blandford-K\"{o}nigl-type conical jet; \citealt{blandford1979}).
A viscous disk without irradiation (MCD model) predicts smaller
values, $\alpha \approx -0.3$ at the flat part between the blackbody
peak frequencies emitted from the outermost and innermost radii, 
and $\alpha=-2$ below it (Rayleigh-Jeans regime).
An irradiated disk predicts $ \alpha \sim -1.2$ (e.g., \citealt{hynes2005}),
although it depends on disk geometry such as the outer disk radius
($R_{\rm out}$ in {\tt diskir}) and irradiation efficiency ($f_{\rm out}$) 
(see e.g., \citealt{shidatsu2011}).
On the basis of the SED fitting results for Epoch (c), we conclude
that the contribution from the irradiated disk is likely not a dominant source
of the optical/UV flux.

To reinforce the interpretation that this component is a jet origin,
  we compare the optical-UV SED in Epoch (f) with the radio flux density at 6 GHz observed with the Very Large Array (VLA) on the same day \citep{2021ApJ...907...34S}. Figure~\ref{fig:unabsorbed_SED_radio} plots the radio to UV SED, where we extrapolate the power-law component determined by the optical-UV SED to lower frequencies. As noticed, the radio flux is consistent with the extrapolation within the uncertainty in the spectral slope. This supports our interpretation that this component is synchrotron radiation from the jets.
Within the wavelength coverage of our SED in the optical-UV region, we
detect neither a power-law spectral break (corresponding to the optically thin to thick transition frequency) nor a high energy cutoff
in the the synchrotron emission (corresponding to the maximum energy
of non-thermal electrons in the jets) and have only an lower limit for it ($>8\times 10^{-3}$ keV).

Figure~\ref{fig:lumin_ratio} plots the X-ray luminosity ($L_{\rm X}$)
versus the optical luminosity of the jet component ($L_{\rm opt}^{\rm
  jet}$). We also add the data point observed in a brighter LHS in the
initial outburst, taken from Figure~9 in \citet{sdt18}.  As noticed, a
clear positive correlation is found, which can be represented by
$L_{\rm opt}^{\rm jet} \propto {L_{\rm X}}^{0.4}$\noprint{xxx}.
The
correlation is similar to what is found between the X-ray and radio
luminosities in the LHS of BHXBs (e.g., \citealt{2013MNRAS.428.2500C}),
and is consistent with the jet interpretation \citep{Russell2006}.
We note that similar correlations are expected also from X-ray irradiated disks
(e.g., \citealt{vanParadijs1994}),
although this model is not supported by the SED fitting (Section~\ref{subsec:SEDana}).

In this paragraph, we focus on the results of Epoch (c)
obtained with the {\tt bknpower*highecut + diskir} model, where the
parameters of {\tt diskir} are well constrained thanks to the
good quality X-ray spectrum covering the 0.4--10 keV range.
The best-fit parameters are consistent with those
  reported by \citet{ozbey2022} except for the innermost disk
  temperature and the disk radius parameter.
In our best-fit SED, the multi-color disk (MCD) component from the standard
disk is dominant at the Extreme UV band, with the innermost temperature ($T_{\rm in}$) of $3.1\times10^{-2}$\noprint{xxx} keV in Epoch (c). 
The X-ray spectrum, which shows no evidence for a soft excess, gives a tight upper limit for $T_{\rm in}$ 
($<0.4$ keV). 
We must make caveats, however, that the best-fit SED we obtain
might be model dependent since the peak wavelength of the MCD
component is not covered by the observed data. 
Bearing that in mind, we discuss implications from the best-fit 
MCD parameters below. The innermost radius is estimated to be
$2.6\times 10^3$\noprint{xxx} km from the normalization of the MCD 
component, suggesting that the standard disk is truncated at 
\noprint{xxx}$2.5 \times 10^2 r_{\rm g}$. Such a truncated disk is considered to be a general
feature of accretion flow in the LHS of BHXBs (e.g., \cite{2009ApJ...707L..87T, 2011PASJ...63S.785S, 2013PASJ...65...80Y}). 
It is noteworthy that the truncation radius is about 10\noprint{xxx} times larger than that reported by \citet{sdt18}  
($\sim 24 r_{\rm g}$) in the LHS when the X-ray luminosity was 
about 30\noprint{xxx} times brighter than our Epoch (c). 
This suggests 
that the innermost radius of a truncated disk
increases with decreasing mass accretion rate in the LHS.

\subsubsection{Period III}
\label{subsec:Disc_SEDPeriod3}

As noticed from Figure~\ref{fig:unabsorbed_SED}, the SED in Epoch (g) (Period III) shows
different features from those in Periods~I and II. A sharp peak is
observed in the optical band, and the flux ratio between the optical
and X-ray bands is much larger than those in the previous periods.
The
analysis of the SED in Epoch (g) (Period III) was performed in
\citet{yoshitake2022},
but for completeness we present shortly the main points here.
Synchro-cyclotron radiation by 
hot electrons is dominant in the optical band, whereas 
the X-ray emission is produced by thermal Bremsstrahlung 
and Comptonization of the synchro-cyclotron photons by 
the same electrons (e.g., \citealt{nar95}, \citealt{man97}).
The relation of the optical jet luminosity and 
the X-ray luminosity in Epoch (g) is also plotted 
in Figure~\ref{fig:lumin_ratio}. As noticed, it generally
follows the relation obtained in the LHS at brighter flux levels.
A slight offset from it is not surprising because the inner disk
structure and X-ray emitting mechanism are not the same 
between the LHS and the ADAF state.

\subsection{Balmer Line Profiles}

The profiles of Balmer emission and absorption lines in the optical
spectra (Figure~\ref{fig:seimei_optSED}) give important clues to
understand the disk structure. 
In general, absorption
lines are formed from the surface of an accretion disk with normal
temperature gradient in vertical directions to the disk plane (i.e.,
decreasing temperature with height), whereas emission lines are
produced from that with ``inverted'' temperature structure \citep{2002ApJ...581.1297J,Hiroi2009}. In X-ray
binaries, such temperature inversion can be produced by strong
irradiation of X-ray/UV emission from the innermost region.

As an alternative possibility, a disk wind, outflowing gas launched from the disk surface, can produce a P-Cygni profile,
  characterized by blueshifted absorption and broad emission
  features. Typical velocities of the absorption by disk winds in
  BHXBs are about a few thousands of km s$^{-1}$ (e.g.,
  \citealt{MunozDarias2016}). In fact, such features were detected
  in optical-near infrared spectra of MAXI J1820 in the LHS
  \citep{MunozDarias2019} and in the HSS \citep{Sanchez2020}
  during the main outburst, although they were very shallow. As
  mentioned in Section~\ref{subsec:seimei_ana}, we are not able to detect clear P-Cygni
  profiles in our {\it Seimei} spectra taken during the rebrightening
  phases. We do not rule out the presence of high-velocity disk winds,
  however, as detection of such features would require a very high
  signal-to-noise ratio spectra.  In the following, we assume that the
  main absorption and line features are not produced by a disk wind.

\begin{figure}[htbp]
  \begin{center}
     \includegraphics[width=80mm]{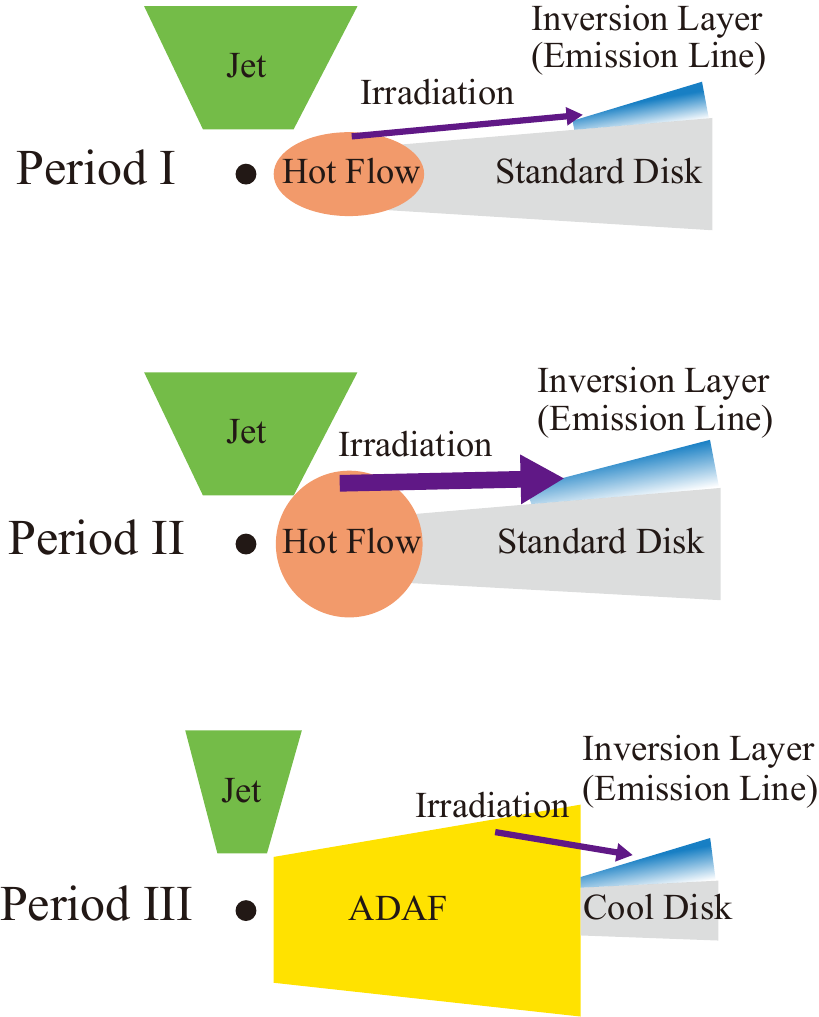} 
  \end{center}
  \caption{
   schematic picture of the accretion disk 
  in the three periods.}
  \label{fig:seimeiSED}
\end{figure}

\subsubsection{Periods I and II}

The line width constrains the inner radius at which the line is
emitted or absorbed, by assuming Keplerian motion in the accretion
disk, as summarized in Table~\ref{tab:SEDseimei}. Here we focus on the results of H$\beta$, because the emission and absorption profiles are well separated (Figure~\ref{fig:seimei_optSED}).
As mentioned above, although the optical spectral observations with
{\it Seimei} were performed in the first and third rebrightening phases, it
is possible to combine the results because of the similarities of the
light curves. We find that the inner radius of the H$\beta$
emitting region slightly decreased from
$\sim 2\times 10^5 r_{\rm g}$ in Period I (7 days
before the X-ray peak) to
$\sim 1\times 10^5 r_{\rm g}$ in Period II (17 days after it).
This result can be interpreted that
the irradiated part of the accretion disk was extended toward inner
radii in Period II. It is consistent with the decrease in the
equivalent width (EW) of the absorption line in Period II because the area
of the unirradiated part (i.e., with normal temperature gradient) had
to shrink accordingly.
A double-peaked H$\beta$ emission profile is noticeable in Period
II, whose peak-to-peak velocity provides a model-independent estimate
of the outer radius of the line-forming region.
The half separation of the peaks is estimated to be $8.5\pm 1.5$~\AA,\noprint{**} 
which corresponds to
$ (3 \pm 1) \times 10^5 r_{\rm g}$ for $i=70^\circ$. This is consistent with
the Roche lobe size ($3 \times 10^5 r_{\rm g}$), which is
adopted as $R_{\rm out}$ in our {\tt diskline} fitting (Section 3.3).

Broad Balmer absorption plus narrow emission features are often
observed in BHXBs during outbursts (e.g., GRO J0422+32,
\citealt{Casares1995}; GRO J1655--40, \citealt{soria2000}; XTE~J1118+480,
\citealt{Dubus2001}; XTE J1859+226, \citealt{Zurita2002}; MAXI J1807+132,
\citealt{Jimenez-Ibarra2019}). The absorption lines are thought to arise in
the accretion disk at smaller radii (as evidenced by their higher
velocities) than the emission lines. In our case, estimated HWZI
values of H$\beta$ absorption
indicate the innermost radii of $\sim (0.6-4) \times 10^3 r_{\rm g}$ (Table~\ref{tab:SEDseimei}). 
\citet{Dubus2001} suggested that the broad absorption lines were 
formed by any combination of a low X-ray luminosity,
a hard X-ray spectrum, or a low inclination (see
their Section 4.1). In our case (MAXI J1820 in the rebrightening phase), 
the first two conditions are satisfied.

Using our best-fit models (Figure~\ref{fig:seimei_optSED}), we calculate
  the equivalent widths (EWs) of H$\alpha$ and H$\beta$ emission lines
  with respect to the ``pseudo'' continua (i.e., those not including
  absorption features) to be $6.1\pm0.2$ \AA~and $3.2\pm0.7$ \AA~(Period~I)
  and $12.3\pm0.2$ \AA~and $3.5\pm0.3$ \AA~(Period~II), respectively. These
  values are within the range of those in the LHS during the main
  outburst reported by \citet{sai2021}, which showed large variability (by
  a factor of $\sim$3 within $<$10 days). Recalling that the jet
  component was dominant in the optical band in Periods I and II, the
  EWs with respect to the ``disk'' components are estimated to be
  2.5--5 times larger than the above values, on the basis of the
  SED fitting result of Epoch (c) with $f_{\rm out}$ free. According to a
  Monte-Carlo radiative transfer simulation of an irradiated accretion
  disk in the HSS of MAXI J1820 by \citet{kolijonen2023}, an emission line
  of H$\alpha$ with a few tens of \AA\ can be produced in the
  reprocessed component by disk atmosphere. Considering various
  uncertainties in the observations (e.g., time variability, modelling
  of the line profiles) and in theoretical calculations (e.g.,
  differences in the SED and disk structure between the LHS and HSS),
  we infer that our results do not contradict with the theoretical predictions. 
  
\subsubsection{Period III}

In Period III, only emission lines are detected in the H$\alpha$ and
H$\beta$ regions (see also \cite{yoshitake2022}, 
who reported a detailed analysis of the H$\alpha$ profile). 
Its width indicates that the line 
emitting region is similar to that in Period~II 
($\sim 10^5 r_{\rm g}$). \citet{yoshitake2022} 
interpret that it is likely to be 
produced from an optically-thick disk in the ``cool'' mode.
The absence of absorption features suggests that an optically-thick
disk in the ``hot'' mode is not present in this state. This is consistent with our SED model, where no MCD component is included.

\subsection{Evolution of Accretion Disk Structure}

Interpreting the results on the SED and Balmer line profiles, we
illustrate our view on the evolution of accretion disk structure
during the rebrightening phase in Figure~\ref{fig:seimeiSED}. 
Note that jets emitting synchrotron radiation in the radio-to-UV band are always present in the system, which are omitted in the figure. In Periods I (rising phase) and II (slowly decaying phase), the overall disk structure is similar to each other. 
The standard disk is truncated before reaching the ISCO, probably at $\sim 100 r_{\rm g}$ around the peak flux. Below the
truncation radius, it turns into a ``hot flow'', which emits 
X-rays by Comptonizing seed photons from the truncated disk. 
In Period~I, hot corona is less developed compared with 
Period II probably because the accretion flow in the innermost region does not reach equilibrium yet. 
This explains the small X-ray to optical luminosity ratios in
Period~I. Also, X-ray irradiation to the outer disk is limited at
large radii in Period~I probably because of geometrical effects (e.g., small scale height of the corona). 
In Period~III (dim phase), the standard disk in the hot mode no longer exists and the whole region at $\lesssim 10^5 r_{\rm g}$ is replaced by ADAF. The UV and X-ray emission from the ADAF irradiates the outer ``cool'' disk at $\gtrsim 10^5 r_{\rm g}$, 
producing Balmer emission lines. 
We repeat our caveats, however, that some of the interpretations rely on specific SED modeling under the limited coverage of the multi-wavelength data. 
Future systematic multi-wavelength observations covering the
whole outburst/rebrightening phase are always important to 
reach robust conclusions on the evolution of the disk structure 
in BHXBs, 
and the history of mass accretion rate through the disk.
The mechanism of rebrightening phenomena often observed in BHXBs and
cataclysmic variables still remains unsolved (e.g. \citealt{Kato1998};
\citealt{Hameury2000}; \citealt{Osaki2001}; \citealt{Meyer2015}).
Those observations will be a key to clarify the rebrightening mechanism.

\section{Conclusion}

In this work, we have analyzed the broadband (near infrared-optical-UV
and X-ray) SED and optical spectra of MAXI J1820 observed in the three
rebrightening phases in 2019 and 2020 following the initial outburst
in 2018. We find that the rebrightening phase can be classified to 3
periods: (Period I) the ``rising'' phase before the X-ray flux peak
when $F_{\rm opt}/F_{\rm X}$ continuously decreased, (Period II)
``decaying'' phase when $F_{\rm opt}/F_{\rm X}$ continuously
increased, and (Period III) ``dim'' phase when $F_{\rm opt}/F_{\rm X}$
was roughly constant. The main conclusions are summarized as below.
\begin{enumerate}

\item In Periods~I and II, MAXI J1820 was always in the LHS, where the
  X-ray spectrum in the 0.4--10 keV band was approximated by a power
  law with a photon index of $\approx1.6$.
  The near-IR/optical/UV SED was dominated by a power-law 
  with a photon index of $0.7\sim1.4$, which smoothly connects to the simultaneous radio flux
  available in the literature in one epoch. We interpret that
  that it originates from synchrotron emission from the jets.

\item The H$\alpha$ and H$\beta$ profiles in Periods I and II show
  broad absorption and narrower emission features. The absorption
  features are centered close to the rest-frame energy, suggesting
  that they are not produced by a disk wind. The line widths indicates
  that the inner radius of an irradiated disk slightly decreased from Period I
  ($\sim 2\times10^5 r_{\rm g}$) to Period II ($\sim 1\times10^5 r_{\rm g}$),
  implying evolution of the hot corona geometry.

\item In Period III, we confirm Yoshitake et al. (2022) that the SED
  can be reproduced by an advection-dominant accretion flow (ADAF)
  model with jet emission. The Balmer line profiles indicate that an
  irradiated cool disk was present at $r>\sim 10^5 r_{\rm g}$.
  
\end{enumerate}

\section*{Acknowledgments}\fontsize{8}{11}\selectfont
We are very grateful to the anonymous referee for providing many
insightful comments to improve the quality of the paper.
This research has made use of data 
and software provided by the High Energy Astrophysics 
Science Archive Research Center (HEASARC), which is 
a service of the Astrophysics Science Division at 
NASA/GSFC. This work made use of public data from 
the {\swift} data archive. We acknowledge the use 
of observations from the Las Cumbres Observatory global 
telescope network. 
Part of this work was financially supported 
by Grants-in-Aid for Scientific Research 19K14762 (MS) 
from the Ministry of Education, Culture, Sports, Science 
and Technology (MEXT) of Japan, and partially 
supported by the Optical and Infrared Synergetic Telescopes 
for Education and Research (OISTER) program funded by the 
MEXT of Japan.
\par
\bibliographystyle{apj}
\bibliography{main}{}

\begin{thebibliography}{}
\expandafter\ifx\csname natexlab\endcsname\relax\def\natexlab#1{#1}\fi

\bibitem[{{Baglio} {et~al.}(2018){Baglio}, {Russell}, \& {Lewis}}]{2018ATel11418....1B}
{Baglio}, M.~C., {Russell}, D.~M., \& {Lewis}, F. 2018, The Astronomer's Telegram, 11418, 1

\bibitem[{{Barden}(1995)}]{bar95}
{Barden}, S.~C., ed. 1995, Society of Photo-Optical Instrumentation Engineers (SPIE) Conference Series, Vol. 2476, {Fiber Optics in Astronomical Applications}, ed. S.~C. {Barden}, 56--67

\bibitem[{{Begelman} {et~al.}(1983){Begelman}, {McKee}, \& {Shields}}]{begelman1983}
{Begelman}, M.~C., {McKee}, C.~F., \& {Shields}, G.~A. 1983, \apj, 271, 70

\bibitem[{{Bertin} \& {Arnouts}(1996)}]{Bertin1996}
{Bertin}, E., \& {Arnouts}, S. 1996, \aaps, 117, 393

\bibitem[{{Blandford} \& {K{\"o}nigl}(1979)}]{blandford1979}
{Blandford}, R.~D., \& {K{\"o}nigl}, A. 1979, \apj, 232, 34

\bibitem[{{Boffin}(2001)}]{2001LNP...573...69B}
{Boffin}, H.~M.~J. 2001, in Astrotomography, Indirect Imaging Methods in Observational Astronomy, ed. H.~M.~J. {Boffin}, D.~{Steeghs}, \& J.~{Cuypers}, Vol. 573 (Springer), 69

\bibitem[{{Burrows} {et~al.}(2005){Burrows}, {Hill}, {Nousek}, {Kennea}, {Wells}, {Osborne}, {Abbey}, {Beardmore}, {Mukerjee}, {Short}, {Chincarini}, {Campana}, {Citterio}, {Moretti}, {Pagani}, {Tagliaferri}, {Giommi}, {Capalbi}, {Tamburelli}, {Angelini}, {Cusumano}, {Br{\"a}uninger}, {Burkert}, \& {Hartner}}]{bur05}
{Burrows}, D.~N., {Hill}, J.~E., {Nousek}, J.~A., {et~al.} 2005, SSRv, 120, 165

\bibitem[{{Casares} {et~al.}(1995){Casares}, {Martin}, {Charles}, {Martin}, {Rebolo}, {Harlaftis}, \& {Castro-Tirado}}]{Casares1995}
{Casares}, J., {Martin}, A.~C., {Charles}, P.~A., {et~al.} 1995, \mnras, 276, L35

\bibitem[{{Chambers} {et~al.}(2016){Chambers}, {Magnier}, {Metcalfe}, {Flewelling}, {Huber}, {Waters}, {Denneau}, {Draper}, {Farrow}, {Finkbeiner}, {Holmberg}, {Koppenhoefer}, {Price}, {Rest}, {Saglia}, {Schlafly}, {Smartt}, {Sweeney}, {Wainscoat}, {Burgett}, {Chastel}, {Grav}, {Heasley}, {Hodapp}, {Jedicke}, {Kaiser}, {Kudritzki}, {Luppino}, {Lupton}, {Monet}, {Morgan}, {Onaka}, {Shiao}, {Stubbs}, {Tonry}, {White}, {Ba{\~n}ados}, {Bell}, {Bender}, {Bernard}, {Boegner}, {Boffi}, {Botticella}, {Calamida}, {Casertano}, {Chen}, {Chen}, {Cole}, {Deacon}, {Frenk}, {Fitzsimmons}, {Gezari}, {Gibbs}, {Goessl}, {Goggia}, {Gourgue}, {Goldman}, {Grant}, {Grebel}, {Hambly}, {Hasinger}, {Heavens}, {Heckman}, {Henderson}, {Henning}, {Holman}, {Hopp}, {Ip}, {Isani}, {Jackson}, {Keyes}, {Koekemoer}, {Kotak}, {Le}, {Liska}, {Long}, {Lucey}, {Liu}, {Martin}, {Masci}, {McLean}, {Mindel}, {Misra}, {Morganson}, {Murphy}, {Obaika}, {Narayan}, {Nieto-Santisteban}, {Norberg}, {Peacock}, {Pier}, {Postman}, {Primak}, {Rae}, {Rai},
  {Riess}, {Riffeser}, {Rix}, {R{\"o}ser}, {Russel}, {Rutz}, {Schilbach}, {Schultz}, {Scolnic}, {Strolger}, {Szalay}, {Seitz}, {Small}, {Smith}, {Soderblom}, {Taylor}, {Thomson}, {Taylor}, {Thakar}, {Thiel}, {Thilker}, {Unger}, {Urata}, {Valenti}, {Wagner}, {Walder}, {Walter}, {Watters}, {Werner}, {Wood-Vasey}, \& {Wyse}}]{2016arX161205560C}
{Chambers}, K.~C., {Magnier}, E.~A., {Metcalfe}, N., {et~al.} 2016, arXiv e-prints, arXiv:1612.05560

\bibitem[{{Corbel} {et~al.}(2013){Corbel}, {Coriat}, {Brocksopp}, {Tzioumis}, {Fender}, {Tomsick}, {Buxton}, \& {Bailyn}}]{2013MNRAS.428.2500C}
{Corbel}, S., {Coriat}, M., {Brocksopp}, C., {et~al.} 2013, \mnras, 428, 2500

\bibitem[{Craine(1994)}]{bar94}
Craine, D. L. C. E.~R., ed. 1994, Society of Photo-Optical Instrumentation Engineers (SPIE) Conference Series, Vol. 2198, {Instrumentation in Astronomy VIII}, ed. D.~L. C. E.~R. Craine, 87--97

\bibitem[{Craine(2010)}]{krt10}
---. 2010, Society of Photo-Optical Instrumentation Engineers (SPIE) Conference Series, Vol. 7733, {Ground-based and Airborne Telescopes III}, ed. D.~L. C. E.~R. Craine

\bibitem[{{C{\'u}neo} {et~al.}(2020){C{\'u}neo}, {Alabarta}, {Zhang}, {Altamirano}, {M{\'e}ndez}, {Armas Padilla}, {Remillard}, {Homan}, {Steiner}, {Combi}, {Mu{\~n}oz-Darias}, {Gendreau}, {Arzoumanian}, {Stevens}, {Loewenstein}, {Tombesi}, {Bult}, {Fabian}, {Buisson}, {Neilsen}, \& {Basak}}]{cuneo2020}
{C{\'u}neo}, V.~A., {Alabarta}, K., {Zhang}, L., {et~al.} 2020, \mnras, 496, 1001

\bibitem[{{Cutri} {et~al.}(2003){Cutri}, {Skrutskie}, {van Dyk}, {Beichman}, {Carpenter}, {Chester}, {Cambresy}, {Evans}, {Fowler}, {Gizis}, {Howard}, {Huchra}, {Jarrett}, {Kopan}, {Kirkpatrick}, {Light}, {Marsh}, {McCallon}, {Schneider}, {Stiening}, {Sykes}, {Weinberg}, {Wheaton}, {Wheelock}, \& {Zacarias}}]{2003tmc..book.....C}
{Cutri}, R.~M., {Skrutskie}, M.~F., {van Dyk}, S., {et~al.} 2003, {2MASS All Sky Catalog of point sources.}

\bibitem[{{Done} {et~al.}(2007){Done}, {Gierli{\'n}ski}, \& {Kubota}}]{don07}
{Done}, C., {Gierli{\'n}ski}, M., \& {Kubota}, A. 2007, A\&A~Rv, 15, 1

\bibitem[{{Dubus} {et~al.}(2001){Dubus}, {Kim}, {Menou}, {Szkody}, \& {Bowen}}]{Dubus2001}
{Dubus}, G., {Kim}, R. S.~J., {Menou}, K., {Szkody}, P., \& {Bowen}, D.~V. 2001, \apj, 553, 307

\bibitem[{{Ebisawa} {et~al.}(1993){Ebisawa}, {Makino}, {Mitsuda}, {Belloni}, {Cowley}, {Schmidtke}, \& {Treves}}]{ebs93}
{Ebisawa}, K., {Makino}, F., {Mitsuda}, K., {et~al.} 1993, ApJ, 403, 684

\bibitem[{{Echibur{\'u}-Trujillo} {et~al.}(2023){Echibur{\'u}-Trujillo}, {Tetarenko}, {Haggard}, {Russell}, {Koljonen}, {Bahramian}, {Wang}, {Bremer}, {Bright}, {Casella}, {Russell}, {Altamirano}, {Baglio}, {Belloni}, {Ceccobello}, {Corbel}, {Diaz Trigo}, {Maitra}, {Gabuya}, {Gallo}, {Heinz}, {Homan}, {Kara}, {K{\"o}rding}, {Lewis}, {Lucchini}, {Markoff}, {Migliari}, {Miller-Jones}, {Rodriguez}, {Saikia}, {Sarazin}, {Shahbaz}, {Sivakoff}, {Soria}, {Testa}, {Tetarenko}, \& {Tudose}}]{Echiburu-Trujillo2023}
{Echibur{\'u}-Trujillo}, C., {Tetarenko}, A.~J., {Haggard}, D., {et~al.} 2023, arXiv e-prints, arXiv:2311.11523

\bibitem[{{Fabian} {et~al.}(1989){Fabian}, {Rees}, {Stella}, \& {White}}]{fabian1989}
{Fabian}, A.~C., {Rees}, M.~J., {Stella}, L., \& {White}, N.~E. 1989, \mnras, 238, 729

\bibitem[{{Gandhi} {et~al.}(2019){Gandhi}, {Rao}, {Johnson}, {Paice}, \& {Maccarone}}]{gan19}
{Gandhi}, P., {Rao}, A., {Johnson}, M. A.~C., {Paice}, J.~A., \& {Maccarone}, T.~J. 2019, MNRAS, 485, 2642

\bibitem[{{Gierli{\'n}ski} {et~al.}(2008){Gierli{\'n}ski}, {Done}, \& {Page}}]{gie08}
{Gierli{\'n}ski}, M., {Done}, C., \& {Page}, K. 2008, \mnras, 388, 753

\bibitem[{{Gierli{\'n}ski} {et~al.}(2009){Gierli{\'n}ski}, {Done}, \& {Page}}]{gie09}
---. 2009, \mnras, 392, 1106

\bibitem[{{Hameury} {et~al.}(2000){Hameury}, {Lasota}, \& {Warner}}]{Hameury2000}
{Hameury}, J.-M., {Lasota}, J.-P., \& {Warner}, B. 2000, \aap, 353, 244

\bibitem[{{HI4PI Collaboration} {et~al.}(2016){HI4PI Collaboration}, {Ben Bekhti}, {Fl{\"o}er}, {Keller}, {Kerp}, {Lenz}, {Winkel}, {Bailin}, {Calabretta}, {Dedes}, {Ford}, {Gibson}, {Haud}, {Janowiecki}, {Kalberla}, {Lockman}, {McClure-Griffiths}, {Murphy}, {Nakanishi}, {Pisano}, \& {Staveley-Smith}}]{2016AA...594A.116H}
{HI4PI Collaboration}, {Ben Bekhti}, N., {Fl{\"o}er}, L., {et~al.} 2016, \aap, 594, A116

\bibitem[{{Hiroi} {et~al.}(2009){Hiroi}, {Moritani}, {Nogami}, {Imada}, {Hashimoto}, {Ueda}, {Soejima}, {Kinugasa}, {Honda}, {Narusawa}, {Sakamoto}, {Iizuka}, {Matsuda}, {Naito}, {Iijima}, \& {Fujii}}]{Hiroi2009}
{Hiroi}, K., {Moritani}, Y., {Nogami}, D., {et~al.} 2009, \pasj, 61, 697

\bibitem[{{Horne} \& {Marsh}(1986)}]{horne1986}
{Horne}, K., \& {Marsh}, T.~R. 1986, \mnras, 218, 761

\bibitem[{{Hynes}(2005)}]{hynes2005}
{Hynes}, R.~I. 2005, \apj, 623, 1026

\bibitem[{{Ishiguro} {et~al.}(2011){Ishiguro}, {Takahashi}, {Zenno}, {Tokimasa}, \& {Kuroda}}]{nic2011}
{Ishiguro}, M., {Takahashi}, J., {Zenno}, T., {Tokimasa}, N., \& {Kuroda}, T. 2011, Annu. Rep. Nishi-Harima Astron. Obs., 21, 13

\bibitem[{{Jimenez-Garate} {et~al.}(2002){Jimenez-Garate}, {Raymond}, \& {Liedahl}}]{2002ApJ...581.1297J}
{Jimenez-Garate}, M.~A., {Raymond}, J.~C., \& {Liedahl}, D.~A. 2002, \apj, 581, 1297

\bibitem[{{Jim{\'e}nez-Ibarra} {et~al.}(2019){Jim{\'e}nez-Ibarra}, {Mu{\~n}oz-Darias}, {Armas Padilla}, {Russell}, {Casares}, {Torres}, {Mata S{\'a}nchez}, {Jonker}, \& {Lewis}}]{Jimenez-Ibarra2019}
{Jim{\'e}nez-Ibarra}, F., {Mu{\~n}oz-Darias}, T., {Armas Padilla}, M., {et~al.} 2019, \mnras, 484, 2078

\bibitem[{{Kato} {et~al.}(1998){Kato}, {Nogami}, {Baba}, \& {Matsumoto}}]{Kato1998}
{Kato}, T., {Nogami}, D., {Baba}, H., \& {Matsumoto}, K. 1998, in Astronomical Society of the Pacific Conference Series, Vol. 137, Wild Stars in the Old West, ed. S.~{Howell}, E.~{Kuulkers}, \& C.~{Woodward}, 9

\bibitem[{{Kawamuro} {et~al.}(2018){Kawamuro}, {Negoro}, {Yoneyama}, {Ueno}, {Tomida}, {Ishikawa}, {Sugawara}, {Isobe}, {Shimomukai}, {Mihara}, {Sugizaki}, {Nakahira}, {Iwakiri}, {Yatabe}, {Takao}, {Matsuoka}, {Kawai}, {Sugita}, {Yoshii}, {Tachibana}, {Harita}, {Morita}, {Yoshida}, {Sakamoto}, {Serino}, {Kawakubo}, {Kitaoka}, {Hashimoto}, {Tsunemi}, {Nakajima}, {Kawase}, {Sakamaki}, {Maruyama}, {Ueda}, {Hori}, {Tanimoto}, {Oda}, {Morita}, {Yamada}, {Tsuboi}, {Nakamura}, {Sasaki}, {Kawai}, {Sato}, {Yamauchi}, {Hanyu}, {Hidaka}, {Yamaoka}, \& {Shidatsu}}]{kwm18}
{Kawamuro}, T., {Negoro}, H., {Yoneyama}, T., {et~al.} 2018, The Astronomer's Telegram, 11399, 1

\bibitem[{{Koljonen} {et~al.}(2023){Koljonen}, {Long}, {Matthews}, \& {Knigge}}]{kolijonen2023}
{Koljonen}, K.~I.~I., {Long}, K.~S., {Matthews}, J.~H., \& {Knigge}, C. 2023, \mnras, 521, 4190

\bibitem[{{Kotani} {et~al.}(2005){Kotani}, {Kawai}, {Yanagisawa}, {Watanabe}, {Arimoto}, {Fukushima}, {Hattori}, {Inata}, {Izumiura}, {Kataoka}, {Koyano}, {Kubota}, {Kuroda}, {Mori}, {Nagayama}, {Ohta}, {Okada}, {Okita}, {Sato}, {Serino}, {Shimizu}, {Shimokawabe}, {Suzuki}, {Toda}, {Ushiyama}, {Yatsu}, {Yoshida}, \& {Yoshida}}]{kot05}
{Kotani}, T., {Kawai}, N., {Yanagisawa}, K., {et~al.} 2005, Nuovo Cimento C Geophysics Space Physics C, 28, 755

\bibitem[{{Kron}(1980)}]{kron1980}
{Kron}, R.~G. 1980, \apjs, 43, 305

\bibitem[{{Lasker} {et~al.}(2008){Lasker}, {Lattanzi}, {McLean}, {Bucciarelli}, {Drimmel}, {Garcia}, {Greene}, {Guglielmetti}, {Hanley}, {Hawkins}, {Laidler}, {Loomis}, {Meakes}, {Mignani}, {Morbidelli}, {Morrison}, {Pannunzio}, {Rosenberg}, {Sarasso}, {Smart}, {Spagna}, {Sturch}, {Volpicelli}, {White}, {Wolfe}, \& {Zacchei}}]{2008AJ....136..735L}
{Lasker}, B.~M., {Lattanzi}, M.~G., {McLean}, B.~J., {et~al.} 2008, \aj, 136, 735

\bibitem[{{Lasota}(2001)}]{lasota2001}
{Lasota}, J.-P. 2001, New Astronomy Reviews, 45, 449

\bibitem[{{Maccarone}(2003)}]{2003AA...409..697M}
{Maccarone}, T.~J. 2003, \aap, 409, 697

\bibitem[{{Makishima} {et~al.}(2008){Makishima}, {Takahashi}, {Yamada}, {Done}, {Kubota}, {Dotani}, {Ebisawa}, {Itoh}, {Kitamoto}, {Negoro}, {Ueda}, \& {Yamaoka}}]{mks08}
{Makishima}, K., {Takahashi}, H., {Yamada}, S., {et~al.} 2008, PASJ, 60, 585

\bibitem[{{Manmoto} {et~al.}(1997){Manmoto}, {Mineshige}, \& {Kusunose}}]{man97}
{Manmoto}, T., {Mineshige}, S., \& {Kusunose}, M. 1997, \apj, 489, 791

\bibitem[{{Matsubayashi} {et~al.}(2019){Matsubayashi}, {Ohta}, {Iwamuro}, {Iwata}, {Kambe}, {Tsutsui}, {Izumiura}, {Yoshida}, \& {Hattori}}]{mtb19}
{Matsubayashi}, K., {Ohta}, K., {Iwamuro}, F., {et~al.} 2019, PASJ, 71, 102

\bibitem[{Matsuoka {et~al.}(2009)Matsuoka, Kawasaki, Ueno, Tomida, Kohama, Suzuki, Adachi, Ishikawa, Mihara, Sugizaki, Isobe, Nakagawa, Tsunemi, Miyata, Kawai, Kataoka, Morii, Yoshida, Negoro, Nakajima, Ueda, Chujo, Yamaoka, Yamazaki, Nakahira, You, Ishiwata, Miyoshi, Eguchi, Hiroi, Katayama, \& Ebisawa}]{mat09}
Matsuoka, M., Kawasaki, K., Ueno, S., {et~al.} 2009, Publications of the Astronomical Society of Japan, 61, 999

\bibitem[{{Meyer} \& {Meyer-Hofmeister}(2015)}]{Meyer2015}
{Meyer}, F., \& {Meyer-Hofmeister}, E. 2015, \pasj, 67, 52

\bibitem[{{Mineshige} \& {Wheeler}(1989)}]{mineshige1989}
{Mineshige}, S., \& {Wheeler}, J.~C. 1989, \apj, 343, 241

\bibitem[{{Mu{\~n}oz-Darias} {et~al.}(2016){Mu{\~n}oz-Darias}, {Casares}, {Mata S{\'a}nchez}, {Fender}, {Armas Padilla}, {Linares}, {Ponti}, {Charles}, {Mooley}, \& {Rodriguez}}]{MunozDarias2016}
{Mu{\~n}oz-Darias}, T., {Casares}, J., {Mata S{\'a}nchez}, D., {et~al.} 2016, \nat, 534, 75

\bibitem[{{Mu{\~n}oz-Darias} {et~al.}(2017){Mu{\~n}oz-Darias}, {Casares}, {Mata S{\'a}nchez}, {Fender}, {Armas Padilla}, {Mooley}, {Hardy}, {Charles}, {Ponti}, {Motta}, {Dhillon}, {Gandhi}, {Jim{\'e}nez-Ibarra}, {Butterley}, {Carey}, {Grainge}, {Hickish}, {Littlefair}, {Perrott}, {Razavi-Ghods}, {Rumsey}, {Scaife}, {Scott}, {Titterington}, \& {Wilson}}]{MunozDarias2017}
---. 2017, \mnras, 465, L124

\bibitem[{{Mu{\~n}oz-Darias} {et~al.}(2019){Mu{\~n}oz-Darias}, {Jim{\'e}nez-Ibarra}, {Panizo-Espinar}, {Casares}, {Mata S{\'a}nchez}, {Ponti}, {Fender}, {Buckley}, {Garnavich}, {Torres}, {Armas Padilla}, {Charles}, {Corral-Santana}, {Kajava}, {Kotze}, {Littlefield}, {S{\'a}nchez-Sierras}, {Steeghs}, \& {Thomas}}]{MunozDarias2019}
{Mu{\~n}oz-Darias}, T., {Jim{\'e}nez-Ibarra}, F., {Panizo-Espinar}, G., {et~al.} 2019, \apjl, 879, L4

\bibitem[{{Narayan} \& {Yi}(1995)}]{nar95}
{Narayan}, R., \& {Yi}, I. 1995, \apj, 452, 710

\bibitem[{{Oasa} {et~al.}(2020){Oasa}, {Ushioda}, {Shibata}, {Seino}, {Kino}, \& {Akitaya}}]{sacraMusashi2020}
{Oasa}, Y., {Ushioda}, K., {Shibata}, Y., {et~al.} 2020, in Ground-based and Airborne Instrumentation for Astronomy VIII, Vol. 11447, International Society for Optics and Photonics, 114475Z

\bibitem[{{Osaki} {et~al.}(2001){Osaki}, {Meyer}, \& {Meyer-Hofmeister}}]{Osaki2001}
{Osaki}, Y., {Meyer}, F., \& {Meyer-Hofmeister}, E. 2001, \aap, 370, 488

\bibitem[{{{\"O}zbey Arabac{\i}} {et~al.}(2022){{\"O}zbey Arabac{\i}}, {Kalemci}, {Din{\c{c}}er}, {Bailyn}, {Altamirano}, \& {Ak}}]{ozbey2022}
{{\"O}zbey Arabac{\i}}, M., {Kalemci}, E., {Din{\c{c}}er}, T., {et~al.} 2022, \mnras, 514, 3894

\bibitem[{{Paczy{\'n}ski}(1971)}]{pat71}
{Paczy{\'n}ski}, B. 1971, \araa, 9, 183

\bibitem[{{Rodi} {et~al.}(2021){Rodi}, {Tramacere}, {Onori}, {Bruni}, {S{\`a}nchez-Fern{\`a}ndez}, {Fiocchi}, {Natalucci}, \& {Ubertini}}]{2021ApJ...910...21R}
{Rodi}, J., {Tramacere}, A., {Onori}, F., {et~al.} 2021, \apj, 910, 21

\bibitem[{{Russell} {et~al.}(2006){Russell}, {Fender}, {Hynes}, {Brocksopp}, {Homan}, {Jonker}, \& {Buxton}}]{Russell2006}
{Russell}, D.~M., {Fender}, R.~P., {Hynes}, R.~I., {et~al.} 2006, \mnras, 371, 1334

\bibitem[{{Russell} {et~al.}(2010){Russell}, {Maitra}, {Dunn}, \& {Markoff}}]{Russell2010}
{Russell}, D.~M., {Maitra}, D., {Dunn}, R.~J.~H., \& {Markoff}, S. 2010, \mnras, 405, 1759

\bibitem[{{Sai} {et~al.}(2021){Sai}, {Wang}, {Wu}, {Lin}, {Feng}, {Zhang}, {Li}, {Zhang}, {Mo}, {Sun}, {Ehgamberdiev}, {Mirzaqulov}, {Rui}, {Lin}, {Zhao}, {Lin}, {Zhang}, {Zhang}, {Zhao}, {Li}, {Xiang}, {Wang}, \& {Wu}}]{sai2021}
{Sai}, H., {Wang}, X., {Wu}, J., {et~al.} 2021, \mnras, 504, 4226

\bibitem[{{Saikia} {et~al.}(2023){Saikia}, {Russell}, {Pirbhoy}, {Baglio}, {Bramich}, {Alabarta}, {Lewis}, \& {Charles}}]{saikia2023}
{Saikia}, P., {Russell}, D.~M., {Pirbhoy}, S.~F., {et~al.} 2023, \apj, 949, 104

\bibitem[{{S{\'a}nchez-Sierras} \& {Mu{\~n}oz-Darias}(2020)}]{Sanchez2020}
{S{\'a}nchez-Sierras}, J., \& {Mu{\~n}oz-Darias}, T. 2020, \aap, 640, L3

\bibitem[{{Shakura} \& {Sunyaev}(1973)}]{sha73}
{Shakura}, N.~I., \& {Sunyaev}, R.~A. 1973, \aap, 500, 33

\bibitem[{{Shaw} {et~al.}(2021{\natexlab{a}}){Shaw}, {Plotkin}, {Miller-Jones}, {Homan}, {Gallo}, {Russell}, {Tomsick}, {Kaaret}, {Corbel}, {Espinasse}, \& {Bright}}]{sha21}
{Shaw}, A.~W., {Plotkin}, R.~M., {Miller-Jones}, J.~C.~A., {et~al.} 2021{\natexlab{a}}, \apj, 907, 34

\bibitem[{{Shaw} {et~al.}(2021{\natexlab{b}}){Shaw}, {Plotkin}, {Miller-Jones}, {Homan}, {Gallo}, {Russell}, {Tomsick}, {Kaaret}, {Corbel}, {Espinasse}, \& {Bright}}]{2021ApJ...907...34S}
---. 2021{\natexlab{b}}, \apj, 907, 34

\bibitem[{{Shidatsu} {et~al.}(2019){Shidatsu}, {Nakahira}, {Murata}, {Adachi}, {Kawai}, {Ueda}, \& {Negoro}}]{sdt19}
{Shidatsu}, M., {Nakahira}, S., {Murata}, K.~L., {et~al.} 2019, ApJ, 874, 183

\bibitem[{{Shidatsu} {et~al.}(2011{\natexlab{a}}){Shidatsu}, {Ueda}, {Tazaki}, {Yoshikawa}, {Nagayama}, {Nagata}, {Oi}, {Yamaoka}, {Takahashi}, {Kubota}, {Cottam}, {Remillard}, \& {Negoro}}]{shidatsu2011}
{Shidatsu}, M., {Ueda}, Y., {Tazaki}, F., {et~al.} 2011{\natexlab{a}}, \pasj, 63, S785

\bibitem[{{Shidatsu} {et~al.}(2011{\natexlab{b}}){Shidatsu}, {Ueda}, {Tazaki}, {Yoshikawa}, {Nagayama}, {Nagata}, {Oi}, {Yamaoka}, {Takahashi}, {Kubota}, {Cottam}, {Remillard}, \& {Negoro}}]{2011PASJ...63S.785S}
---. 2011{\natexlab{b}}, \pasj, 63, S785

\bibitem[{{Shidatsu} {et~al.}(2018){Shidatsu}, {Nakahira}, {Yamada}, {Kawamuro}, {Ueda}, {Negoro}, {Murata}, {Itoh}, {Tachibana}, {Adachi}, {Yatsu}, {Kawai}, {Hanayama}, {Horiuchi}, {Akitaya}, {Saito}, {Takayama}, {Ohshima}, {Katoh}, {Takahashi}, {Nagayama}, {Yamanaka}, {Kawabata}, {Nakaoka}, {Takagi}, {Morokuma}, {Morihana}, {Maehara}, \& {Sekiguchi}}]{sdt18}
{Shidatsu}, M., {Nakahira}, S., {Yamada}, S., {et~al.} 2018, ApJ, 868, 54

\bibitem[{{Shimokawabe} {et~al.}(2008){Shimokawabe}, {Kawai}, {Kotani}, {Yatsu}, {Ishimura}, {Vasquez}, {Mori}, {Kudo}, {Yoshida}, {Yanagisawa}, {Nagayama}, {Toda}, {Shimozu}, {Kuroda}, {Watanabe}, {Fukushima}, \& {Mori}}]{mitsumeAkeno2008}
{Shimokawabe}, T., {Kawai}, N., {Kotani}, T., {et~al.} 2008, in American Institute of Physics Conference Series, Vol. 1000, Gamma-ray Bursts 2007, ed. M.~{Galassi}, D.~{Palmer}, \& E.~{Fenimore}, 543--546

\bibitem[{{Soria} {et~al.}(2000){Soria}, {Wu}, \& {Hunstead}}]{soria2000}
{Soria}, R., {Wu}, K., \& {Hunstead}, R.~W. 2000, \apj, 539, 445

\bibitem[{{Stiele} \& {Kong}(2020)}]{stiele_and_kong2020}
{Stiele}, H., \& {Kong}, A.~K.~H. 2020, \apj, 889, 142

\bibitem[{{Takahashi} {et~al.}(2013){Takahashi}, {Zenno}, \& {Ishiguro}}]{nic2013}
{Takahashi}, J., {Zenno}, T., \& {Ishiguro}, M. 2013, Bull. Cent. Astron. Univ. Hyogo, 1, 17–22

\bibitem[{{Tanaka} \& {Shibazaki}(1996)}]{tanaka1996}
{Tanaka}, Y., \& {Shibazaki}, N. 1996, \araa, 34, 607

\bibitem[{{Tetarenko} {et~al.}(2021){Tetarenko}, {Casella}, {Miller-Jones}, {Sivakoff}, {Paice}, {Vincentelli}, {Maccarone}, {Gandhi}, {Dhillon}, {Marsh}, {Russell}, \& {Uttley}}]{2021MNRAS.504.3862T}
{Tetarenko}, A.~J., {Casella}, P., {Miller-Jones}, J.~C.~A., {et~al.} 2021, \mnras, 504, 3862

\bibitem[{{Tetarenko} {et~al.}(2023){Tetarenko}, {Shaw}, \& {Charles}}]{tetarenko2023}
{Tetarenko}, B.~E., {Shaw}, A.~W., \& {Charles}, P.~A. 2023, \mnras, 526, 6284

\bibitem[{{Tetarenko} {et~al.}(2016){Tetarenko}, {Sivakoff}, {Heinke}, \& {Gladstone}}]{tetarenko2016}
{Tetarenko}, B.~E., {Sivakoff}, G.~R., {Heinke}, C.~O., \& {Gladstone}, J.~C. 2016, \apjs, 222, 15

\bibitem[{{Tomsick} {et~al.}(2009){Tomsick}, {Yamaoka}, {Corbel}, {Kaaret}, {Kalemci}, \& {Migliari}}]{2009ApJ...707L..87T}
{Tomsick}, J.~A., {Yamaoka}, K., {Corbel}, S., {et~al.} 2009, \apjl, 707, L87

\bibitem[{{Torres} {et~al.}(2019{\natexlab{a}}){Torres}, {Stefanik}, \& {Latham}}]{tor19}
{Torres}, G., {Stefanik}, R.~P., \& {Latham}, D.~W. 2019{\natexlab{a}}, ApJ, 885, 9

\bibitem[{{Torres} {et~al.}(2020){Torres}, {Casares}, {Jim{\'e}nez-Ibarra}, {{\'A}lvarez-Hern{\'a}ndez}, {Mu{\~n}oz-Darias}, {Armas Padilla}, {Jonker}, \& {Heida}}]{tor20}
{Torres}, M.~A.~P., {Casares}, J., {Jim{\'e}nez-Ibarra}, F., {et~al.} 2020, \apjl, 893, L37

\bibitem[{{Torres} {et~al.}(2019{\natexlab{b}}){Torres}, {Casares}, {Jim{\'e}nez-Ibarra}, {Mu{\~n}oz-Darias}, {Armas Padilla}, {Jonker}, \& {Heida}}]{2019ApJ...882L..21T}
---. 2019{\natexlab{b}}, ApJL, 882, L21

\bibitem[{{Tucker} {et~al.}(2018){Tucker}, {Shappee}, {Holoien}, {Auchettl}, {Strader}, {Stanek}, {Kochanek}, {Bahramian}, {ASAS-SN}, {Dong}, {Prieto}, {Shields}, {Thompson}, {Beacom}, {Chomiuk}, {ATLAS}, {Denneau}, {Flewelling}, {Heinze}, {Smith}, {Stalder}, {Tonry}, {Weiland}, {Rest}, {Huber}, {Rowan}, \& {Dage}}]{tuk18}
{Tucker}, M.~A., {Shappee}, B.~J., {Holoien}, T.~W.~S., {et~al.} 2018, \apjl, 867, L9

\bibitem[{{Ueda} {et~al.}(1998){Ueda}, {Inoue}, {Tanaka}, {Ebisawa}, {Nagase}, {Kotani}, \& {Gehrels}}]{ueda1998}
{Ueda}, Y., {Inoue}, H., {Tanaka}, Y., {et~al.} 1998, \apj, 492, 782

\bibitem[{{Uttley} {et~al.}(2018){Uttley}, {Gendreau}, {Markwardt}, {Strohmayer}, {Bult}, {Arzoumanian}, {Pottschmidt}, {Ray}, {Remillard}, {Pasham}, {Steiner}, {Neilsen}, {Homan}, {Miller}, {Iwakiri}, \& {Fabian}}]{2018ATel11423....1U}
{Uttley}, P., {Gendreau}, K., {Markwardt}, C., {et~al.} 2018, The Astronomer's Telegram, 11423, 1

\bibitem[{{Vahdat Motlagh} {et~al.}(2019){Vahdat Motlagh}, {Kalemci}, \& {Maccarone}}]{2019MNRAS.485.2744V}
{Vahdat Motlagh}, A., {Kalemci}, E., \& {Maccarone}, T.~J. 2019, \mnras, 485, 2744

\bibitem[{{van Paradijs} \& {McClintock}(1994)}]{vanParadijs1994}
{van Paradijs}, J., \& {McClintock}, J.~E. 1994, \aap, 290, 133

\bibitem[{{Wilms} {et~al.}(2000){Wilms}, {Allen}, \& {McCray}}]{2000ApJ...542..914W}
{Wilms}, J., {Allen}, A., \& {McCray}, R. 2000, ApJ, 542, 914

\bibitem[{{Yamada} {et~al.}(2013){Yamada}, {Makishima}, {Done}, {Torii}, {Noda}, \& {Sakurai}}]{2013PASJ...65...80Y}
{Yamada}, S., {Makishima}, K., {Done}, C., {et~al.} 2013, \pasj, 65, 80

\bibitem[{{Yanagisawa} {et~al.}(2010){Yanagisawa}, {Kuroda}, {Yoshida}, {Shimizu}, {Nagayama}, {Toda}, {Ohta}, \& {Kawai}}]{mitsumeOkayama2010}
{Yanagisawa}, K., {Kuroda}, D., {Yoshida}, M., {et~al.} 2010, in American Institute of Physics Conference Series, Vol. 1279, Deciphering the Ancient Universe with Gamma-ray Bursts, ed. N.~{Kawai} \& S.~{Nagataki}, 466--468

\bibitem[{{Yang} {et~al.}(2015){Yang}, {Xie}, {Yuan}, {Zdziarski}, {Gierli{\'n}ski}, {Ho}, \& {Yu}}]{2015MNRAS.447.1692Y}
{Yang}, Q.-X., {Xie}, F.-G., {Yuan}, F., {et~al.} 2015, \mnras, 447, 1692

\bibitem[{{Yatsu} {et~al.}(2007){Yatsu}, {Kawai}, {Shimokawabe}, {Vasquez}, {Ishimura}, {Kotani}, {Yanagisawa}, {Yoshida}, {Nagayama}, {Shimizu}, {Toda}, \& {Kuroda}}]{mitsumeAkeno2007}
{Yatsu}, Y., {Kawai}, N., {Shimokawabe}, T., {et~al.} 2007, Physica E Low-Dimensional Systems and Nanostructures, 40, 434

\bibitem[{{Yoshida}(2005)}]{ysd05}
{Yoshida}, M. 2005, Journal of Korean Astronomical Society, 38, 117

\bibitem[{{Yoshitake} {et~al.}(2022){Yoshitake}, {Shidatsu}, {Ueda}, {Mineshige}, {Murata}, {Adachi}, {Maehara}, {Nogami}, {Negoro}, {Kawai}, {Niwano}, {Hosokawa}, {Saito}, {Oasa}, {Takarada}, {Shigeyoshi}, \& {Oister Collaboration}}]{yoshitake2022}
{Yoshitake}, T., {Shidatsu}, M., {Ueda}, Y., {et~al.} 2022, \pasj, 74, 805

\bibitem[{{Zhang} {et~al.}(2019){Zhang}, {Bernardini}, {Russell}, {Gelfand}, {Lasota}, {Qasim}, {AlMannaei}, {Koljonen}, {Shaw}, {Lewis}, {Tomsick}, {Plotkin}, {Miller-Jones}, {Maitra}, {Homan}, {Charles}, {Kobel}, {Perez}, \& {Doran}}]{zhang2019}
{Zhang}, G.~B., {Bernardini}, F., {Russell}, D.~M., {et~al.} 2019, \apj, 876, 5

\bibitem[{{Zurita} {et~al.}(2002){Zurita}, {S{\'a}nchez-Fern{\'a}ndez}, {Casares}, {Charles}, {Abbott}, {Hakala}, {Rodr{\'\i}guez-Gil}, {Bernabei}, {Piccioni}, {Guarnieri}, {Bartolini}, {Masetti}, {Shahbaz}, {Castro-Tirado}, \& {Henden}}]{Zurita2002}
{Zurita}, C., {S{\'a}nchez-Fern{\'a}ndez}, C., {Casares}, J., {et~al.} 2002, \mnras, 334, 999

\end{thebibliography}



\end{document}